\documentclass[10pt]{article}
\usepackage{grffile}
\usepackage{ctable}

\usepackage{amssymb}
\usepackage{amsmath}
\usepackage{mathrsfs}
\usepackage[dvips]{epsfig}
\usepackage[small]{caption}
\usepackage{graphicx}
\usepackage[all]{xy}

 {\begin{trivlist} \item[]{\bf Proof. }}%
 {\hspace*{\fill}$\rule{.4\baselineskip}{.4\baselineskip}$\end{trivlist}}

\newenvironment{Acknowledgment}%
 {\begin{trivlist}\item[]\textbf{Acknowledgments.}}{\end{trivlist}}

\makeatletter\@addtoreset{figure}{section}\makeatother

\makeatletter \@addtoreset{equation}{section} \makeatother

\newcommand{\R}{\mathbb{R}}
\newcommand{\C}{\mathbb{C}}

\def\Re{\mathop{\mathrm{Re}}}
\def\Im{\mathop{\mathrm{Im}}}

\newcommand{\rmO}{\mathrm{O}}

\newcommand{\rmd}{\mathrm{d}}
\newcommand{\rme}{\mathrm{e}}
\newcommand{\rmi}{\mathrm{i}}

\newcommand{\eps}{\varepsilon}

\makeatletter
\newsavebox{\@brx}
\newcommand{\llangle}[1][]{\savebox{\@brx}{\(\m@th{#1\langle}\)}%
  \mathopen{\copy\@brx\kern-0.5\wd\@brx\usebox{\@brx}}}
\newcommand{\rrangle}[1][]{\savebox{\@brx}{\(\m@th{#1\rangle}\)}%
  \mathclose{\copy\@brx\kern-0.5\wd\@brx\usebox{\@brx}}}
\makeatother

\definecolor{Green}{rgb}{0.,0.4,0.}

\renewcommand{\leq}{\leqslant}
\renewcommand{\geq}{\geqslant}

\newcommand{\Rmnum}[1]{\uppercase\expandafter{\romannumeral #1\relax}}

\def\XXint#1#2#3{{\setbox0=\hbox{$#1{#2#3}{\int}$}
     \vcenter{\hbox{$#2#3$}}\kern-.5\wd0}}

\newfam\bifam
\font\tenbi=cmmib10 scaled \magstep1 \font\sevenbi=cmmib10 at 11pt
\font\fivebi=cmmib10 at 6pt \textfont\bifam = \tenbi
\scriptfont\bifam = \sevenbi \scriptscriptfont\bifam= \fivebi

\ifx\pdfoutput\undefined
   \pdffalse
\else
   \pdfoutput=1
   \pdftrue
\fi
\ifpdf
   \usepackage{graphicx}
   \usepackage{epstopdf}
\else
   \usepackage{graphicx}
\fi

\begin{document}

\begin{center}

{\fontsize{16}{16}\fontfamily{cmr}\fontseries{b}\selectfont{Wavenumber 
selection in coupled transport equations}}\\[0.2in]
Arnd Scheel$\,^1$ and Angela Stevens$\,^2$\\
\textit{\footnotesize 
$\,^1$University of Minnesota, School of Mathematics,   206 Church St. S.E., Minneapolis, MN 55455, USA\\
$\,^2$Universit\"at M\"unster, Fachbereich Mathematik und Informatik, Einsteinstrasse 62,
48149 M\"unster}
\date{\small \today} 
\end{center}

\begin{abstract}
\noindent 
We study mechanisms for wavenumber selection in a minimal model for 
run-and-tumble dynamics. We show that nonlinearity in tumbling rates 
induces the existence of a plethora of traveling- and standing-wave patterns, 
as well as a subtle selection mechanism for the wavenumbers of
spatio-temporally periodic waves. We comment on possible implications for 
rippling patterns observed in colonies of myxobacteria. 
\end{abstract}

%

\vspace*{0.2in}

{\small
{\bf Running head:} {Pattern selection in coupled transport equations}

}
\vspace*{0.2in}

%
%
%
%

\section{Introduction}

Inspired by Turing's \cite{Turing} proposition of pattern formation mechanisms based solely on simple diffusion and reaction mechanisms, there has been a tremendous amount of theoretical and experimental efforts devoted to designing and studying systems that exhibit mechanisms for the selection of spatial structure. Theoretically, the most accessible scenario is the instability of an unpatterned, spatially homogeneous state against perturbations with spatial structure. 
The linearly fastest growing mode then allows for rough 
predictions of wavenumbers in nonlinear systems. Despite the simple 
theoretical appeal of Turing's prediction, it 
took a long time 
to support his theoretical predictions experimentally, see e.g. 
\cite{kepboi,DeKepper}, 
where sustained, self-organized  spatially periodic structures are observed in an open-flow chemical reaction;   
\cite{Schier}, where the roles of the Nodal and Lefty signal
are summarized  
during the blastula stage; 
\cite{SickRTS}, where WNT and DKK signaling is discussed in the
context of hair follicle spacing; and  
\cite{Epstein},  
where Turing's ideas are quantitatively tested in a cellular
chemical system.


It's worth noticing that absence of diffusion in one species out of two, 
does not lead to selection of finite wavenumbers in linear instabilities 
\cite{ermlew,diffode}. 
However, patterns with well-defined wavenumber laws had 
experimentally 
been observed for such systems long
before Turing's prediction \cite{liese-orig,liesegang,liese2}. 
On the other hand,  
systems of two species do not exhibit linear instabilities that select temporal 
oscillations with a finite wavenumber. Such spatio-temporal selections through a fastest growing linearly unstable mode, that 
will be the main theme of the present paper, are possible in reaction-diffusion systems with at least three species, only; see \cite{Turing,erm,ermlew}.
%

Our interest here is in a yet simpler system of run-and-tumble dynamics. 
We think of self-propelled agents { moving} with the 
{ same} speed to either right or left. In addition, agents may 
change orientation, and start { moving} in the opposite 
direction. The probability of this tumbling events is assumed to be a 
pointwise function of the densities of left- and right-moving agents. The 
resulting tumbling rates can be thought of as encoding probabilities of 
encounters between left- and right-moving agents, which in turn induce 
changes of orientation. 

\begin{figure}
\centering
\includegraphics[height=1.7in]{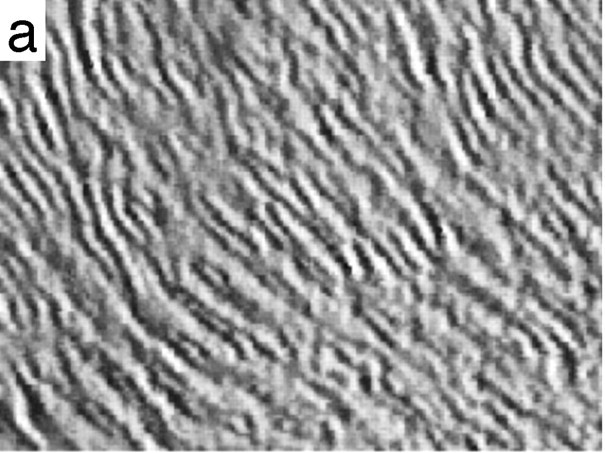}\qquad\qquad
\includegraphics[height=1.7in]{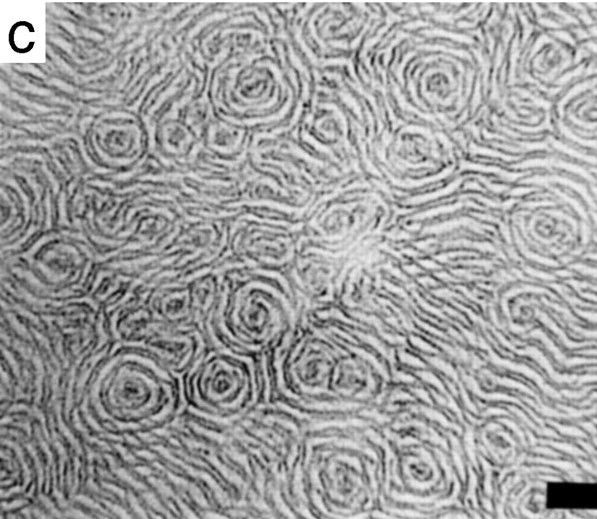}
\caption{Experimentally observed patterns of high and low cell
density in myxobacteria colonies, 
forming planar ripple patterns (left) or concentric circles and spiral 
wave patterns (right); from \cite[Fig. 5]{igoshin} with permission.}
\label{f:exp}
\end{figure}

This caricature picture is motivated by observations of rippling patterns 
in colonies of myxobacteria \cite{Reichenbach65, ShimketsKaiser82, 
SagerKaiser94, WelchKaiser01}. 
During aggregation and starvation induced self-organization 
of myxobacteria an
intriguing rippling pattern can be observed within the colony; 
see Figure \ref{f:exp}.
The ripple crests are oriented approximately perpendicular to the
movement direction of the bacteria. The so-called C-signal, bound
to their cell surface, is transmitted upon end-to-end contact of bacteria,
and increases the reversal probability of individual bacteria. A number of 
mathematical models with different assumption have been discussed
in the literature in order to decribe this rippling behavior, 
cf. \cite{IMWKO, LutscherStevens, BDRB,
AlberJK, BoernerBaer04, IgoshinNeuOster04, igoshin,  AndersonVasiev, SNZO06, 
BDB06, PrimiStevensVelazquez, BGM}.   

All models include active motion and collision-triggered reversal of
myxobacteria. { Continuous,  \cite{BoernerBaer04, BGM, IMWKO,
IgoshinNeuOster04, igoshin, LutscherStevens, PrimiStevensVelazquez}, 
and discrete approaches, \cite{ 
AlberJK, AndersonVasiev, BDB06, BDRB, SNZO06} }  
are discussed. Several of the models assume a refractory period
after collision and/or a reversal triggering internal biochemical
clock to play a role. This is not needed in \cite{LutscherStevens, 
BoernerBaer04}. In \cite{PrimiStevensVelazquez} different cellular states
are suggested instead, i.e. the state of excitation (being able to receive or
send the C-signal), the state when being in contact with counter-migrating
cells, and the state 
of reversing the motors.

Although rippling patterns could numerically be observed in all models, a 
strict argument that the respective assumed mechanisms really do select a 
wavenumber,
as one can see in experiments, could so far only 
be shown for the model in  \cite{PrimiStevensVelazquez}.
As pointed out in \cite{IgoshinNeuOster04, PrimiStevensVelazquez}
any linear stability analysis is not capable to predict a correct
wavelength and wavespeed for the models given in \cite{IMWKO, LutscherStevens}.
In the limit of weak signaling for the model 
in \cite{IMWKO} a Fokker-Planck type of
equation was derived by \cite{BGM} for the reversal-point 
density, that contains a 
source term which is absent in the respective limit given by 
\cite{IgoshinNeuOster04}. For small nonlinearities wave number 
selection could be found. Strengthening the nonlinearity tends to
confine and destroy the patterns through a nonequilibrium phase
transition, reminiscent of destruction of synchronization in the
Kuramoto model. \\
Such an analysis is even less accessible for discrete or individual
based models, i.e. leaving 
the rigorous analysis of the suggested mechanisms
for rippling patterns and defined wavelengths in these 
models still largely open. 
A further crucial control for the suggested rippling mechanisms are 
their capability to also reflect mutant, or dilution-experiments, 
see Section \ref{s:6} for a detailed discussion, and the 
close relation between rippling and aggregation patterns. 
 
Tying in with these considerations and 
since the actual mechanism for motion and tumbling are most definitely 
quite complex, we think of our caricature example as 
trying to exhibit the simplest mechanism that can explain the intriguing 
rippling patterns and also the associated selection of wavenumbers. 

To be precise, we start with the model that was analyzed 
in \cite{LutscherStevens},
\begin{align}
u_t&=+u_x-r(u,v)+r(v,u),\nonumber\\
v_t&=-v_x+r(u,v)-r(v,u),\label{e:cte}
\end{align}
where $u=u(t,x)$ and $v=v(t,x)$ encode the densities of left- and right-moving 
bacteria, respectively, $r(u,v)$ is the rate at which left-moving bacteria 
reverse direction, and, by reflection symmetry, $r(v,u)$ the rate at which 
right-moving bacteria reverse direction. Such
systems do arise as special lower dimensional case from
structured population dynamics models, cf. \cite{PrimiStevensVelazquez,
StevensVelazquez},
\begin{equation}\label{e:astr}
\partial_t U (t,x,c) + V(c) \cdot \partial_x U (t,x,c)
+ \partial_c \left[ K\left ( U(t,x,c) \right) \right] = 0,
\end{equation}
where $\{ c \}$ denotes a set of internal variables which
characterize the state of the considered species, i.e. here
the direction of motion. The most distinctive feature of
the operator $K$ is, that it acts on the density $U$ 
in a local manner. To our knowledge, the classification of pattern 
formation in
such systems is largely open, and one could view our attempts at a 
precise description of the patterns in the drastically simplified 
system \eqref{e:cte} as a first step towards a systematic description of 
patterns in the more complex system \eqref{e:astr}.

At times, we will also allude to a diffusive regularization
of (\ref{e:cte}), 
\begin{align}
u_t&=\eps^2 u_{xx}+u_x-r(u,v)+r(v,u)\nonumber\\
v_t&=\eps^2 v_{xx}-v_x+r(u,v)-r(v,u),\label{e:cted}
\end{align}
with $\eps>0$, small, encoding Brownian fluctuations in addition to 
directed motion. We mostly think of these systems posed on $x\in\R$, but will restrict to periodic boundary conditions when convenient. 

While most of our results give valid predictions for general systems of the form \eqref{e:cte}, we focus in our analysis on the following class of tumbling rates 
\begin{equation}\label{e:tr}
r(u,v)=u\cdot g(v), \qquad g(v)=1+\frac{v^2}{1+\gamma v^2},
\end{equation}
for some $\gamma\geq 0$. Those nonlinearities model a tumbling rate proportional to the number of say right-moving agents, depending in a nonlinear fashion on encounters with oppositely oriented agents. Constants $g(v)\equiv 1$ encode spontaneous tumbling, linear dependence $g(v)=v$ encodes binary collisions, albeit with zero net effect on the dynamics due to the presence of the reverse tumbling, $ug(v)-vg(u)=0$ for $g(v)=v$. Therefore we keep the next simplest term 
$g(v)=v^2$, which encodes triple collisions, and we also allow for a 
Hill-type saturation of the tumbling rate for large densities through the denominator $1+\gamma v^2$. 

The interaction of tumbling and transport appears to be surprisingly 
difficult to characterize. The linear transport equation associated with 
linear tumbling rates $g(v)\equiv \mu$ can be converted to a damped wave 
equation for $\rho=u+v$ using the ``Kac trick'',
\begin{equation}\label{e:kac}
\rho_{tt} + 2\mu \rho_t - \rho_{xx}=0,
\end{equation}
with diffusive long-time dynamics for $\mu>0$. For nonlinear tumbling rates, 
such a simple description does not appear to be possible. One can however 
find criteria that guarantee global existence of solutions, at least in the 
presence of small viscosity $\eps>0$ \eqref{e:cted}; 
see \cite{Hillen97,LutscherStevens}. In fact, invariance of the 
domain $(u,v)\in [0,s]^2$ for \eqref{e:cte} implies $r(w,s)-r(s,w)\leq 0$ 
for all $0\leq w\leq s$, which, for $r(u,v)=ug(v)$ gives 
\[
g(s)w-g(w)s\leq 0,\quad \mbox{for} \ 0<w<s.
\]
For power-law behavior, those conditions imply sublinear growth of $g$, 
motivating to some extent the choice of a Hill-type saturation in 
\eqref{e:tr}. 

Systems \eqref{e:cte} and \eqref{e:cted} exhibit trivial, spatially 
constant equilibrium densities \footnote{Equilibrium here refers to 
an Eulerian, 
density equilibrium --- the agents are perpetually moving and tumbling.} 
when $r(u,v)=r(v,u)$, which is satisfied for 
$u\equiv v$ \emph{(symmetric states)}, but possibly also along curves 
where $u\not\equiv v$ \emph{(asymmetric states)}. Following Turing's ideas
\cite{Turing}, one can then ask for the type of patterns which may emerge from 
instabilities of such uniform densities, striving to find a simple 
explanation for the occurance of the ripples shown in Figure \ref{f:exp}. 
It was noted in \cite{LutscherStevens, PrimiStevensVelazquez}
that the linearization at symetric states does \emph{not} predict finite 
wavelength patterns as fastest growing modes of the linearization in this 
simple two-species models. In fact, at the onset of instability, \emph{all} 
spatial wavenumbers become simultaneously neutrally stable, with eigenvalues 
on the imaginary axis, and past onset spatially homogeneous perturbations 
exhibit the fastest growth rate; see also Section \ref{s:2}, below. 
``Turing instabilities'', where the first instability occurs for a 
wavenumber $0<k_*<\infty$ arise only when more complexity is allowed, 
for instance the introduction
of different stages of right and left moving bacteria, 
$u_j,v_j$, $j \geq 2$; see \cite{PrimiStevensVelazquez, StevensVelazquez}.

Our main results here can be informally summarized as follows: \emph{
\begin{enumerate}
\item linear growth favors wavenumbers $k_\mathrm{lin}=0$ or $k_\mathrm{lin}=\infty$, that is, linear instabilities \emph{do not select} finite wavenumbers from white-noise perturbations;
\item localized perturbations of asymmetric states may generate traveling waves with a selected non-zero wavenumber $k_\mathrm{loc}$, that is,  instabilities \emph{do select} finite wavenumbers from shot-noise perturbations;
\item localized perturbations of symmetric states result in the creation of 
asymmetric states and subsequent evolution of traveling and standing waves, with nonzero wavenumber $k_\mathrm{loc}$, that is, localized instabilities \emph{eventually do select} finite wavenumbers from shot noise perturbations.
\end{enumerate}
}
 The key insight is that localized perturbations result in a spatio-temporal 
\emph{spreading} of perturbations. The resulting invasion process is 
oscillatory in nature with a well-defined spreading speed and finite temporal 
frequency. In other words, 
\emph{oscillatory invasion selects spatial wavenumbers}. 

Such spatio-temporal selection mechanisms are also relevant in contexts where the run-and-tumble dynamics are subject to an additional growth mechanism. We demonstrate this here in an oversimplified scenario, where growth is induced by simply depositing agents into the system at locations $x=\pm ct$. 
Most importantly, the selection mechanisms referred to above are not induced by diffusion. Indeed, linear growth from white noise may favor a finite 
wavenumber $0<k(\eps)<\infty$ in \eqref{e:cted}, but $k(\eps)\to 0$ or 
$k(\eps)\to\infty$ for $\eps\to 0$, for these kinds of perturbations.  
For shot noise perturbations, diffusion plays a subordinate role, and the selected wavenumber $k_\mathrm{loc}(\eps)$ is continuous in $\eps$ with nonzero limit at $\eps=0$. 

These main findings are summarily illustrated below in Figure \ref{f:7} 
(no wavenumber selection from white noise perturbations), Figure  \ref{f:4} 
(wavenumber selection from local perturbations),  Figure \ref{f:6} 
(wavenumber selection from shot noise perturbations), 
Figure \ref{f:8} (wavenumber selection from localized perturbations of 
symmetric states), and Figure \ref{f:9} (wavenumber selection through growth).

\textbf{Outline:} We discuss kinetics and instabilities of spatially uniform distributions in Section \ref{s:2}. Section \ref{s:3} contains existence and stability analysis of nonlinear traveling- and standing-wave patterns. We briefly review linear pointwise growth theory in Section \ref{s:4} and apply the theory to our specific example. Section \ref{s:5} demonstrates the validity of these predictions in several contexts, including the above mentioned localized and shot noise perturbations of localized equilibria. We conclude with a discussion and list of open problems, Section \ref{s:6}.

\begin{Acknowledgment}
A. Scheel was partially supported through NSF grants DMS-1612441 
and DMS-1311740, through a DAAD  Faculty Research Visit Grant, WWU Fellowship, 
and a Humboldt Research Award. 
A. Stevens was partially supported by the DFG Excellence Cluster 
Cells in Motion
(CiM).
A. Scheel gratefully acknowledges generous hospitality during his extended research stay at the WWU M\"unster. 
\end{Acknowledgment}

\section{Tumbling kinetics and linear analysis}\label{s:2}

We first discuss dynamics of $x$-independent density profiles, Section \ref{s:2.1} and then calculate stability of these profiles against $x$-dependent perturbations, Section \ref{s:2.2}. 

\subsection{Tumbling kinetics}\label{s:2.1}

The tumbling kinetics 
\begin{align}
u_t&=-r(u,v)+r(v,u)\nonumber\\
v_t&=+r(u,v)-r(v,u),\label{e:tumbl}
\end{align}
can in principle be solved explicitly, exploiting mass conservation 
$u(t)+v(t)\equiv M$ to reduce to a scalar equation 
$u_t=-r(u,M-u)+r(M-u,u)$.  Possibly more elegantly, notice the reflection 
symmetry $(u,v)\mapsto (v,u)$ and 
introduce invariant $\rho=u+v$ and equivariant $m=u-v$ as variables, 
\[
\rho_t=0, \qquad m_t=-2\left(r\left(\frac{\rho+m}{2},\frac{\rho-m}{2}\right)
-r\left(\frac{\rho-m}{2},\frac{\rho+m}{2}\right)\right)
,\]
which in turn can be written as 
\[
\rho_t=0,\qquad m_t=mR\left(m^2,\rho\right),
\]
exploiting the fact that $m_t$ is an odd function in $m$ and viewing $\rho$ 
as a parameter. The symmetric solution branch $m=0$, changes stability when $\partial_1R=0$, or
\begin{equation}\label{e:symstab}
\partial_1 r(u,u)=\partial_2 r(u,u) 
\end{equation}
and a nontrivial branch of solutions with $m\neq 0$ bifurcates in a local pitchfork bifurcation when stability changes. 

\begin{figure}[h]
	\centering
	\includegraphics[width=0.245\linewidth]{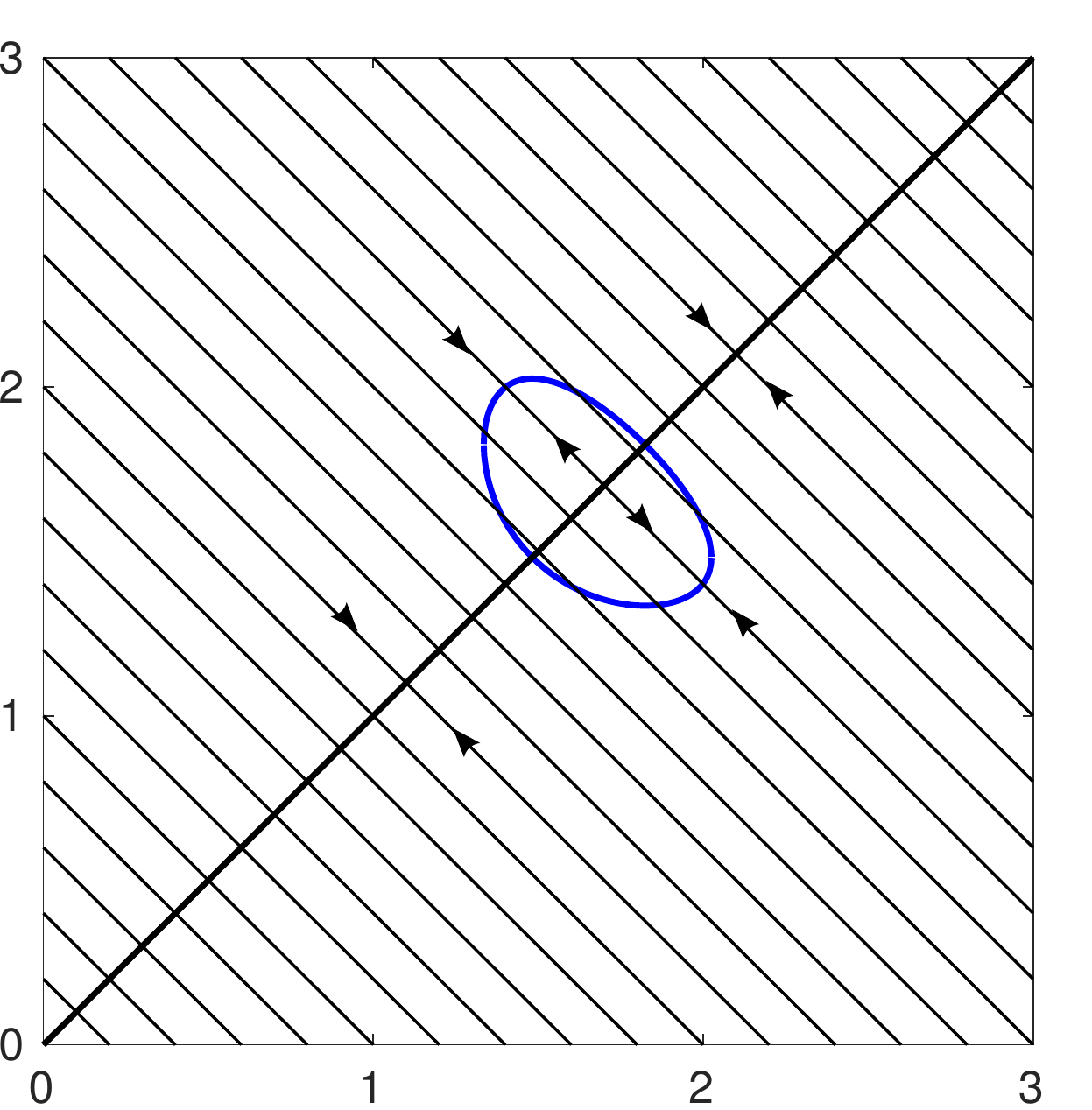}\hfill\includegraphics[width=0.245\linewidth]{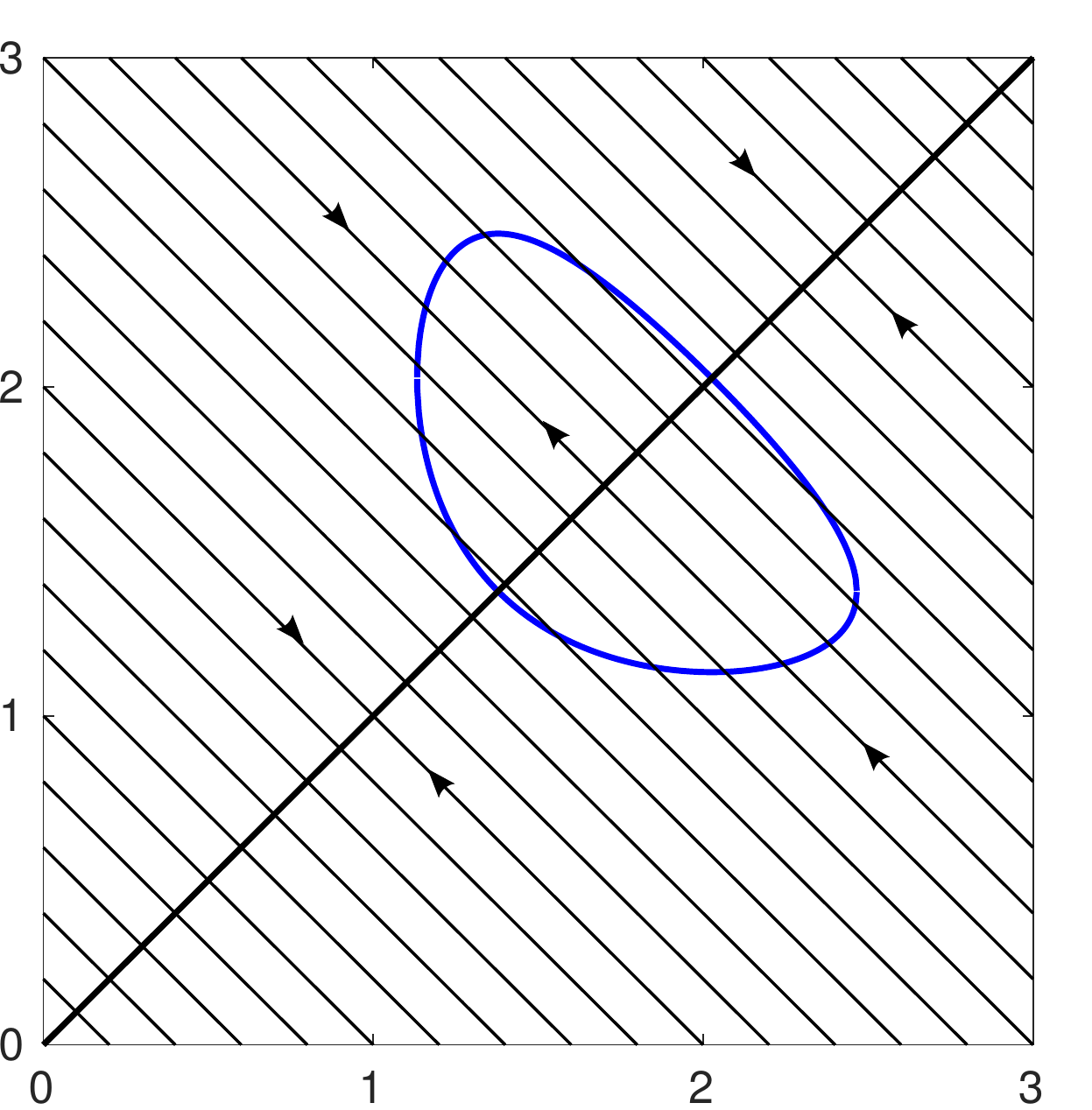}\hfill\includegraphics[width=0.245\linewidth]{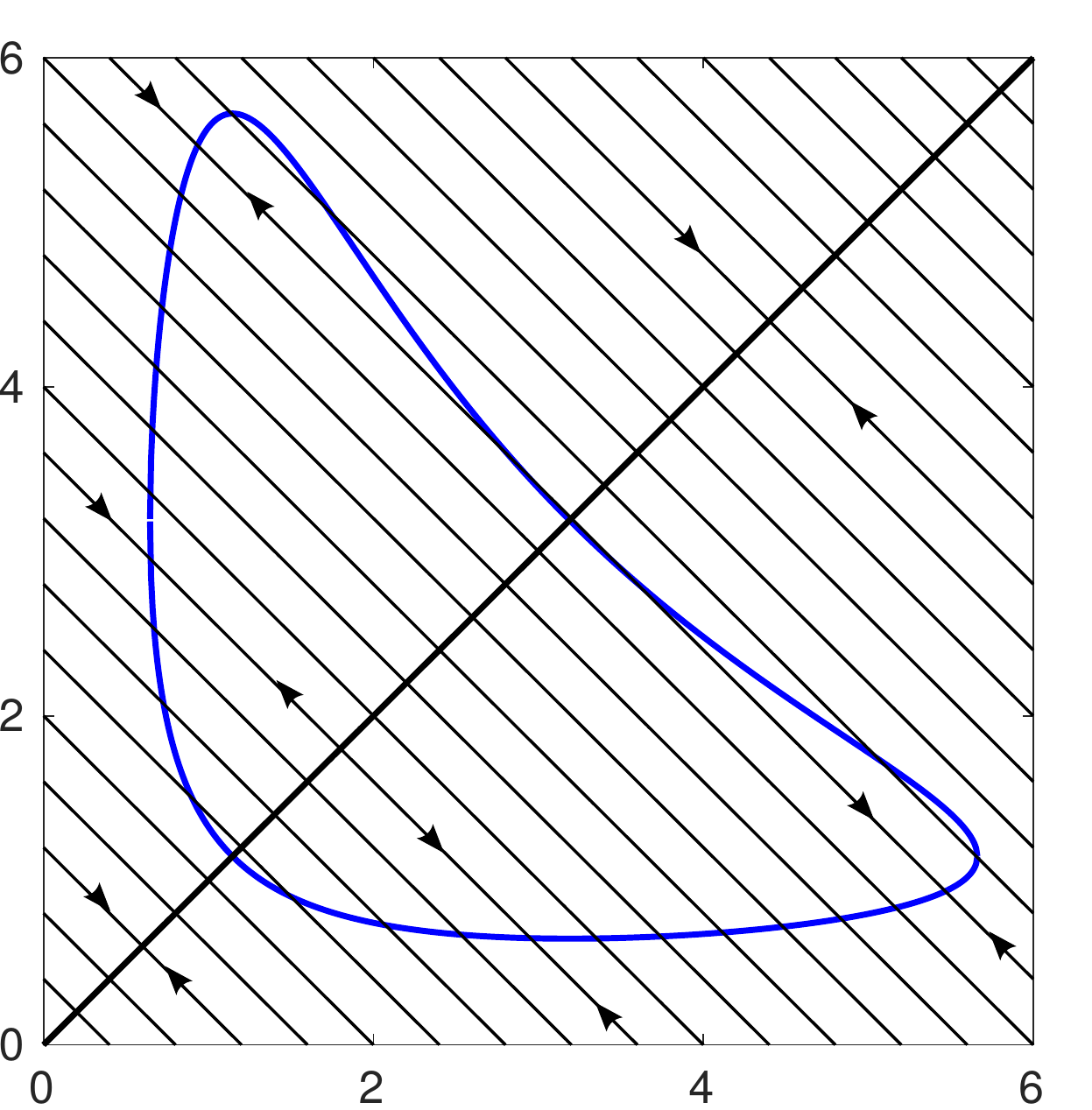}\hfill\includegraphics[width=0.25\linewidth]{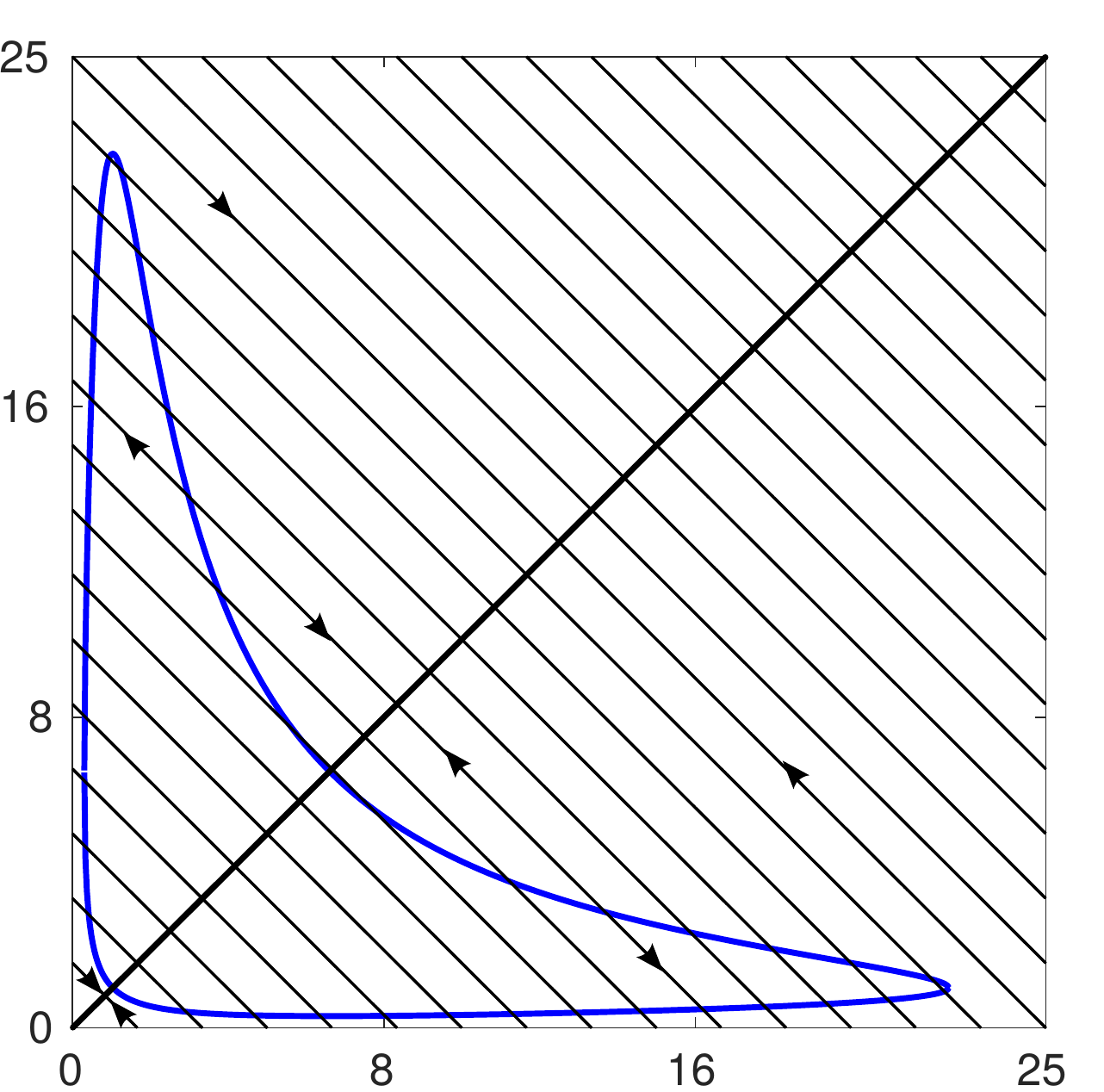}
	\caption{Dynamics of tumbling kinetics \eqref{e:tumbl} { in the phase 
plane, with coordinate axes for $u$ (horizontal) and $v$. Kinetics $r$ are as in \eqref{e:tr} with parameter values $\gamma=0.122,0.115,0.07,0.021$ from left 
to right.} Stability changes correspond to horizontal tangencies of curves of equilibria.}\label{f:1}
\end{figure}

Asymmetric equilibria continue as curves, generically without further bifurcation points. Linear stability of an equilibrium $(u,v)=(u_*,v_*)$ is readily found from the linearized matrix
\[
A_*=\left(\begin{array}{cc}
-\partial_1 r^{uv}+\partial_2 r^{vu} & \partial_1 r^{vu}-\partial_2 r^{uv} \\
\partial_1 r^{uv}-\partial_2 r^{vu} & -\partial_1 r^{vu}+\partial_2 r^{uv} 
\end{array}\right)=\left(\begin{array}{cc}
n_1 & n_2 \\
-n_1 & -n_2 
\end{array}\right)
\]
where the superscripts $uv$ and $vu$ indicate evaluation at $(u_*,v_*)$, respectively, and 
\[
n_1:=-\partial_1 r^{uv}+\partial_2 r^{vu},\qquad n_2:= \partial_1 r^{vu}-\partial_2 r^{uv}.
\]
Note that $n=(n_1,n_2)$ is a normal vector to a curve of equilibria as it is perpendicular to the kernel of $A_*$. Asymmetric equilibria are stable when the nontrivial eigenvalue of $A_*$ is negative, 
\[
\lambda_2:=n_1-n_2<0.
\] 
Bifurcation diagrams for the specific kinetics \eqref{e:tr} are shown in Figure \ref{f:1} and illustrate the relation between orientation of curves of equilibria, that is, of $n$, and stability.
In fact, asymmetric equilibria exist for $\gamma<1/8$ 
for $g$ like in (\ref{e:tr}) and are given explicitly through 
\[
u=\frac{v\pm\sqrt{v^2-4\gamma\left(1+\gamma v^2\right)\left(1+\left(1+\gamma\right)v^2\right)}}{2\gamma\left(1+\left(1+\gamma\right)v^2\right)}
\]
bifurcating from the symmetric  branch at 
\[
u_*=v_*=\sqrt{\frac{1-2\gamma\pm \sqrt{1-8\gamma}}{2\gamma(1+\gamma)}}.
\]
For $\gamma=0$, asymmetric equilibria are always stable for the pure 
tumbling kinetics and lie on the hyperbola $uv=1$.

\subsection{Dispersion relations and linear stability}\label{s:2.2}

\begin{figure}[h!]
	\centering
	\includegraphics[width=\linewidth]{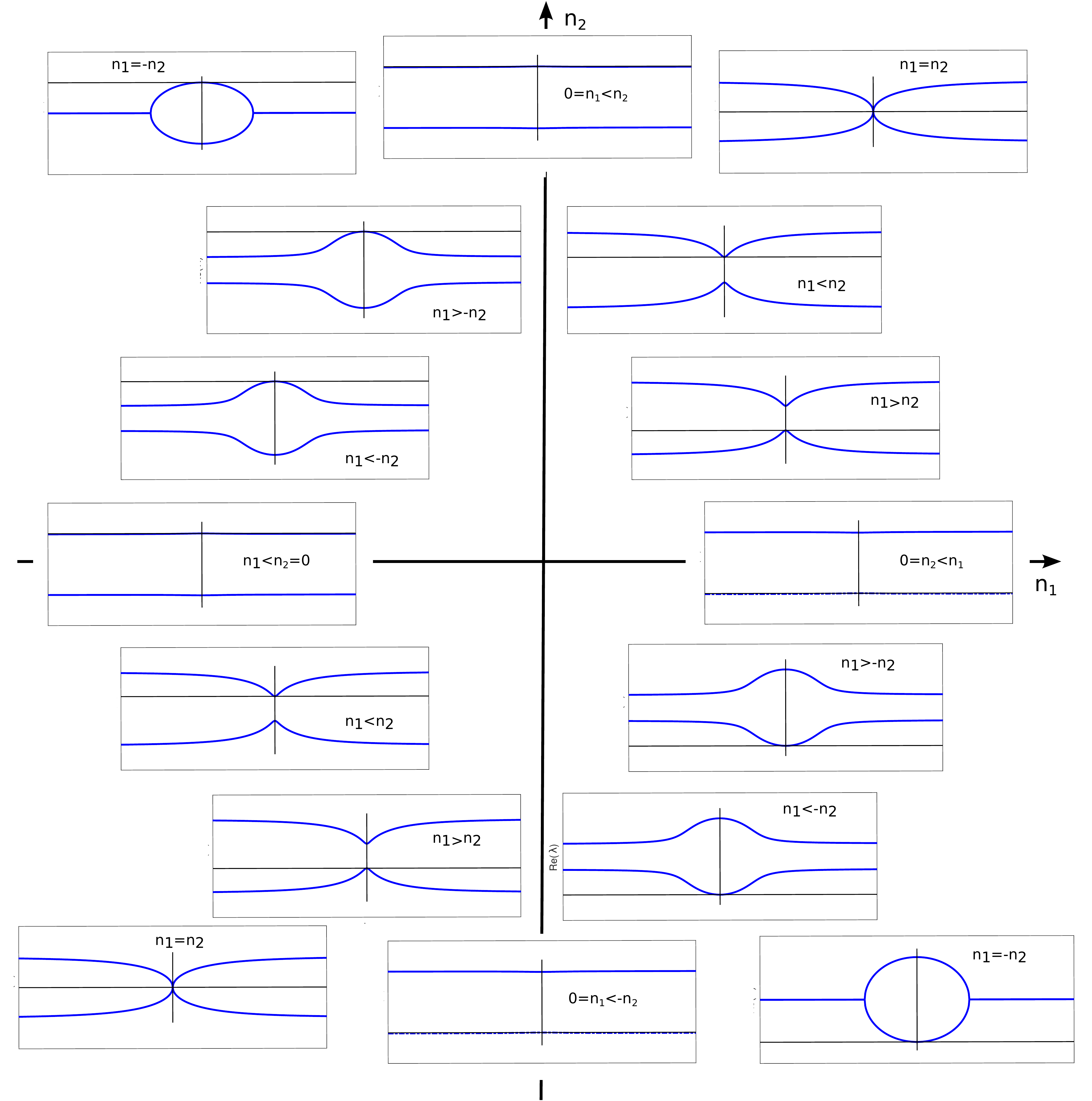}
	\caption{Dispersion relations $\Re\lambda_\pm$ as functions of $k$ from \eqref{e:dislam} for typical choices of $(n_1,n_2)$.}\label{f:2}
\end{figure}
\begin{figure}[h!]
	\centering
	\includegraphics[width=\linewidth]{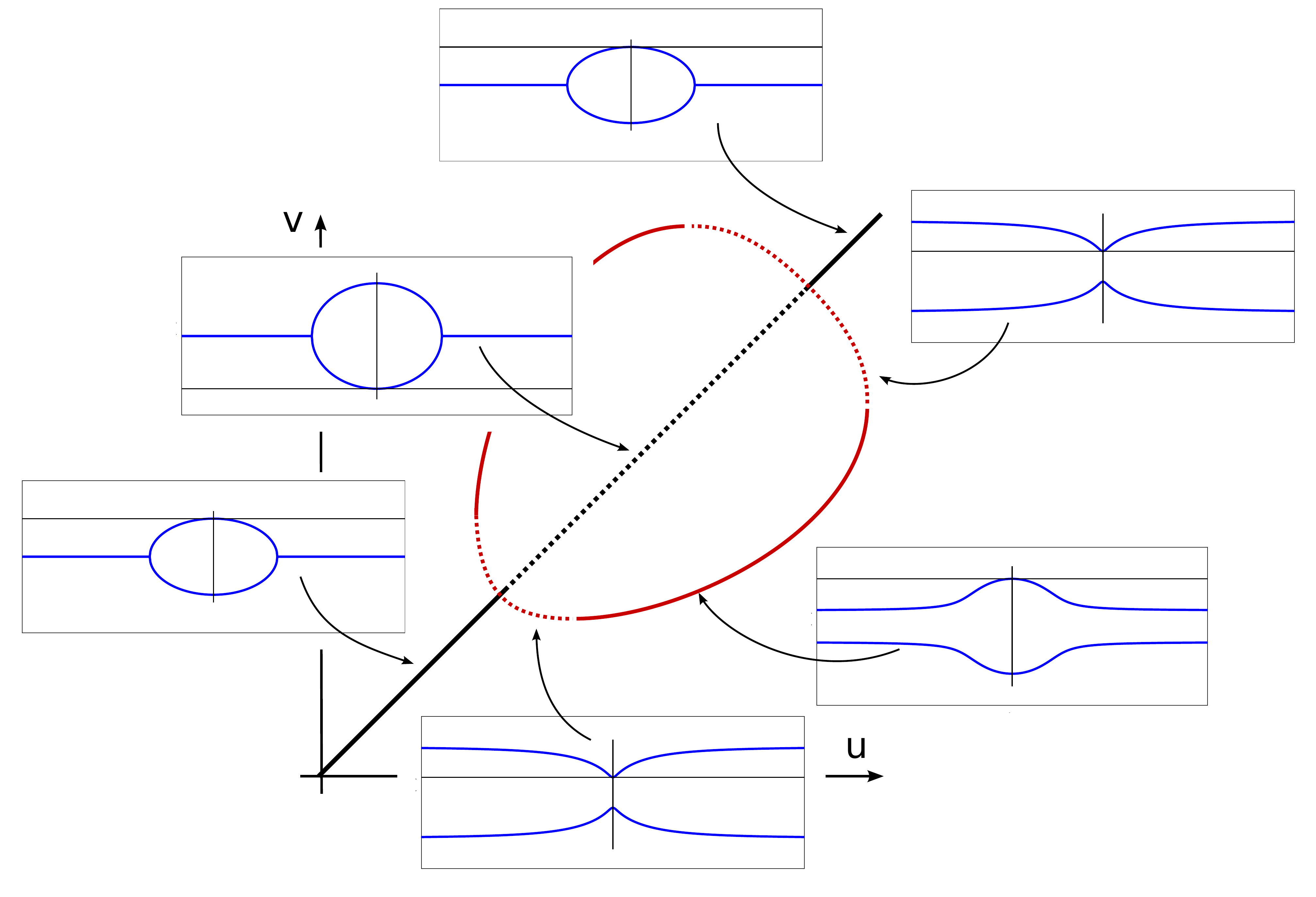}
	\caption{Dispersion relations $\Re\lambda_\pm(k)$ from \eqref{e:dislam} along branches of equilibria. Stability depends on the tangent vectors of asymmetric branches, only. }\label{f:3}
\end{figure}

We now turn to analyzing stability of spatially homogeneous states with respect to $x$-dependent perturbations. After Fourier transformation with 
variable $\nu=\rmi k$, we find the family of matrices 
\[
A_*(\nu)=\left(\begin{array}{cc}
-\partial_1 r^{uv}+\partial_2 r^{vu} +\nu & \partial_1 r^{vu}-\partial_2 r^{uv} \\
\partial_1 r^{uv}-\partial_2 r^{vu} & -\partial_1 r^{vu}+\partial_2 r^{uv} -\nu
\end{array}\right)=\left(\begin{array}{cc}
n_1 +\nu& n_2 \\
-n_1 & -n_2-\nu 
\end{array}\right),
\]
with eigenvalues $\lambda$ found as roots of the dispersion relation, 
\begin{equation}\label{e:dis}
d(\lambda,\nu)=\mathrm{det}\,\left(A_*(\nu)-\lambda\right)=\lambda\left(\lambda-(n_1-n_2)\right)-\nu\left(\nu+(n_1+n_2)\right),
\end{equation}
that is, setting $\nu=\rmi k$,
\begin{equation}\label{e:dislam}
\lambda_\pm(k)=\frac{1}{2}\left(n_1-n_2\pm \sqrt{(n_1-n_2)^2+4\rmi k\left(n_1+n_2+\rmi k\right)}  \right).
\end{equation}
Expanding  the neutral branch, $\lambda_-(0)=0$, at $k=0$, we find
\begin{equation}\label{e:explam}
\lambda_-(k)=-c_\mathrm{g}\rmi k-d_\mathrm{eff} k^2 + \rmO(k^3),\qquad \mbox{with } c_\mathrm{g}=\frac{n_1+n_2}{n_1-n_2}, \  d_\mathrm{eff}=4\frac{n_1 n_2}{(n_1-n_2)^3},
\end{equation}
such that long-wavelength modulations can be thought of as obeying a 
diffusive transport law with transport 
velocity given by the group velocity $c_\mathrm{g}$ and effective viscosity $d_\mathrm{eff}$ nonzero, even at $\eps=0$. One readily finds curves $\Re\lambda_\pm(k)$ as depicted in Figure \ref{f:2}. Stable equilibria have $n_1<0<n_2$. 
Note that $\Re\lambda_\pm$ is maximal for $k=0$ on the symmetric branch, and when $n_1<0<n_2$ or $n_2<0<n_1$ on the asymmetric branch, and for $k=\infty$, 
otherwise. Consequences for stability are illustrated in the bifurcation diagrams 
in Figure \ref{f:3}. 
We remark that the presence of diffusion would cause all branches $\lambda_\pm(k)$ to stabilize eventually, $\Re\lambda_\pm \sim -k^2$ for $|k|$ large. One 
can therefore verify that in the cases where the selected wavenumber is 
$k=\infty$, a small amount of diffusion would induce selection of finite 
wavenumbers, that is $\Re\lambda$ maximal for some $0<k(\eps)<\infty$. 
On the other hand, $k_*(\eps)\to \infty$ as $\eps\to 0$, such that pattern 
selection occurs on the length scale of diffusion
and should therefore not be thought of as induced by the tumbling mechanism. In direct simulations, we also noticed that such patterns were typically subject to coarsening, driving the system eventually to an unpatterned state. 

Comparing with experiments, the linear stability of symmetric states with small 
mass  $(u,v)\equiv m$ agrees with the observed absence of rippling 
patterns in bacterial colonies with small mass densities 
\cite{noripple}.

\section{Nonlinear traveling- and standing-wave patterns}\label{s:3}

We exhibit a large family of traveling- and standing-wave patterns and 
comment on stability. 

\paragraph{Traveling waves: non-characteristic speeds.}
Traveling-wave solutions $u(x-ct),v(x-ct)$ solve the differential equation
\begin{align}
(-1-c)u_\xi&=-r(u,v)+r(v,u)\nonumber\\
(1-c)v_\xi&=+r(u,v)-r(v,u).\label{e:tw}
\end{align}
For $|c|\neq 1$, this is a genuine differential equation, bounded solutions are smooth, with conserved quantity $M_c=(-1-c)u+(1-c)v$. As a consequence, we can reduce to a scalar ODE with heteroclinic orbits as the only possible bounded solutions. Geometrically, the resulting dynamics can immediately be inferred from 
intersecting the bifurcation diagram with lines of constant $M_c$. 

{Since level lines of $M_c$ are not simply anti-diagonals, it is possible for such heteroclinic orbits to connect two equilibria that are both stable for the pure tumbling kinetics. One can however verify that one of the two equilibria is necessarily linearly unstable for the PDE, that is, $\Re\lambda_\pm(k)>0$ for some $k$,  for any $c$ and any choice of tumbling kinetics.}
%

\paragraph{Traveling waves: characteristic speeds.}
 {When $c=1$, the $v$-equation
is algebraic and solutions are not 
necessarily smooth as functions of $\xi$. From the algebraic relation, we find that $r(u,v)=r(v,u)$  for 
all $\xi$, such that $u_\xi=0$ and $u\equiv u_*$ is 
constant, while $v$ may jump arbitrarily between roots of $r(u_*,v)-r(v,u_*)$. As a consequence, traveling waves can be constructed by fixing a value $u_*$ and choosing $v$ arbitrarily for any $\xi$ on branches of the bifurcation diagram, provided that for this particular value of $u=u_*$ there exist multiple equilibrium solutions $v$. The simplest such solution amounts to a front mediating  one vertical jump in the bifurcation diagram.}

\paragraph{Standing and counter-propagating waves.}
We can more generally look for solutions with ``compatible'' values that is, the measurable initial condition takes values on equilibria, only,
\[
u_0(x)\in \{u_*,v_*\},\quad 
v_0(x)\in \{u_*,v_*\},\quad r(u_*,v_*)=r(v_*,u_*), \quad u_*\neq v_*.
\]
We then define $u(t,x)=u_0(x+t)$, $v(t,x)=v_0(x-t)$ and claim that $(u(t,x),v(t,x))$ is a solution to the initial value problem. Indeed, the nonlinearity vanishes on the solution for all times, since $(u,v)$-values are either on the diagonal or at a zero of the combined tumbling rate, and the transport term is accommodated by the shift. We briefly discuss in the appendix some basic existence theory, defining a concept of a solution that allows for local existence and 
uniqueness, but also for discontinuous solutions of this form. 

Summarizing, the particular structure of our system imposes very strong 
restrictions on traveling waves, in particular on smooth traveling waves, 
but allows for a very large family of step-function type wave patterns. 
On those patterns, the nonlinarity vanishes such that dynamics are simply 
left- and right-shifts in the two components, equivalent to the wave equation. 

\paragraph{Stability.} {The arguably most interesting 
 situation
are discontinuous fronts with speed $c=1$ (or $c=-1$), $u\equiv u_*$, $v(x)=v^\pm$ for $\pm x>0$. In the comoving frame of reference, waves solve a transport equation coupled to an ODE, with linearization
\[
u_t = 2u_x + n_1 u + n_2 v,\qquad v_t = -n_1 u - n_2 v,
\]
and the notation from Section \ref{s:2} for derivatives of the tumbling 
rates. The coefficients and $n_j=n_j(x)=n_j^\pm$ for $\pm x>0$  are associated with the linearization at $(u_*,v^\pm)$}. While it would be interesting to prove nonlinear stability in general, we restrict ourselves here to showing that there do not exist unstable linear modes associated with the discontinuity. Therefore, 
we restrict ourselves to $n_1^\pm-n_2^\pm>0$, that is, both asymptotic states 
are stable, and consider the eigenvalue problem
\[
\lambda u = 2u_x + n_1 u + n_2 v,\qquad \lambda v = -n_1 u - n_2 v,
\]
with $\Re\lambda\geq 0$. For $\lambda+n_2^\pm\neq 0$, this equation reduces 
to an ODE for $u$ with no bounded solutions for $\lambda$ to the right of the essential spectrum. In case $\lambda=-n_2^->0$, say, we conclude $n_1<0$ such that $n_1-n_2<0$, and therefore $u=0$ from the second equation. The first equation then also gives $v=0$, hence, by continuity at $x=0$, $u=v=0$ and again $\lambda$ is not an eigenvalue.

\section{Wavenumber selection mechanisms --- linear pointwise growth}\label{s:4}
We review the basic idea of pointwise growth, state the pinched double root criterion, and compute linear predictions in our coupled transport system. 

\paragraph{Pointwise growth.}
As pointed out in the introduction, our main goal here is to emphasize the 
interplay of transport and instability, particularly through its role in 
selecting wavenumbers. The ideas  relate back to  considerations
in plasma physics, where a distinction between 
convective and absolute instabilities was deemed important \cite{briggs,bers}. 
The concepts became more generally relevant in fluid mechanics \cite{HM} and were later 
on discovered to be important for instabilities in dissipative, pattern-forming systems \cite{dee,vS}. 

The key idea is to analyze the growth of spatially localized initial 
conditions in a finite window of observation. For a linear equation 
$u_t=L u$ with constant coefficients, one finds the solution via Laplace transform 
\[
u(t,\cdot)=\frac{1}{2\pi\rmi}\int_\Gamma \rme^{\lambda t}(\lambda - L)^{-1}
u(0,\cdot)\, \rmd \lambda=\frac{1}{2\pi\rmi}\int_\Gamma \rme^{\lambda t}\int_\R G_\lambda(\cdot -y)u(0,y)\,\rmd y\,\rmd  \lambda,
\]
where $\Gamma$ is a contour in the complex half plane to the right of  
singularities of the integrand, and $G_\lambda$ is the Green's function 
associated with the resolvent $(\lambda-L)^{-1}$. In order to find temporal 
asymptotics, one deforms the contour $\Gamma$ exploiting analyticity of the integrand such as to minimize  
the maximum of its real part. This process is limited by the presence of singularities of the integrand in the complex plane. {The key observation is
that $G_\lambda(\xi)$ might be analytic for fixed $\xi$ in regions of the complex $\lambda$-plane even 
though $(\lambda-L)$ is not bounded invertible, in particular not analytic, in those regions. For compactly supported initial data $u(0,y)$, and evaluating $u(t,\cdot)$ at fixed spatial locations, one may then be able to deform the contour of integration in the formulation using the Green's function further and obtain exponential decay although $(\lambda-L)$ may possess singularities in $\Re\lambda>0$.} The simplest example is the operator $\partial_x$ on $L^2$, where $(\lambda-\partial_x)$ is not bounded invertible for $\Re\lambda=0$, but the Green's function $G_\lambda(\xi)=\rme^{\lambda\xi}\chi_{\xi<0}$ is analytic for all $\lambda\in \C$. This reflects the fact that compactly 
supported initial conditions decay to zero in finite time, hence faster than 
any exponential, when observed in a finite window of $x$-values. We refer to 
\cite{da} for a more mathematical recent account of the linear theory and its relation to front invasion problems, but also to \cite{fhs} for limitations of this linear approach. 

\paragraph{The pinched double root criterion.}
The Green's function  $G_\lambda$ in our case can be found through solving the 
constant-coefficient ODE, 
\begin{align*}
u_x&=-\frac{1}{1+c}\left((n_1-\lambda ) u + n_2 v-u_0\delta(x)\right),\\
v_x&=\frac{1}{1-c}\left(-n_1 u -(n_2+\lambda) v-v_0\delta(x)\right),
\end{align*}
where $c$ denotes the speed of the frame of observation. Solutions to the homogeneous equation are exponentials 
$(u^\pm_\lambda,v^\pm_\lambda)\rme^{\nu_\pm(\lambda) x}$, where $(u^\pm_\lambda,v^\pm_\lambda)$ are eigenvectors to the{ coefficient} matrix on 
the right-hand side with eigenvalues $\nu_\pm(\lambda)$. Eigenvectors and eigenvalues, and hence also $G_\lambda$, are analytic as long as $\nu_+\neq \nu_-$. Obstructions to analyticity are therefore \emph{pinched double roots}. In fact, 
the $\nu_\pm(\lambda)$ are roots to the complex extension of the dispersion relation, 
$
d_c(\lambda,\nu):=d(\lambda- c\nu,nu)=0,
$
{ where $d$ was} defined in \eqref{e:dis}. The eigenvalues $\nu_+=\nu_-$ correspond to a double root when in addition $\frac{\rmd}{\rmd \nu} d_c(\lambda ,\nu)=0$. The ``pinching condition'' usually refers to the requirement that $\Re\nu_\pm(\lambda)\to \pm\infty$ as $\Re\lambda\to +\infty$; it is automatically satisfied in our setting as long as $|c|<1$, and never satisfied when $|c|>1$, since in that case both roots $\Re\nu_\pm\to+\infty$ or $\Re\nu_\pm\to-\infty$. This is intuitively clear because information cannot propagate faster than the characteristic speeds; transport is unidirectional in such a frame of reference and the Green's function is analytic for all $\lambda$ just as in the simple example of the operator $\partial_x$, described above.

Double roots $\lambda_*,\nu_*$, that is, solutions to 
\begin{equation}\label{e:dr}
d_c(\lambda,\nu)=0,\qquad \partial_\nu d_c(\lambda,\nu)=0,
\end{equation}
therefore yield a \emph{pointwise growth rate} and a growth frequency. One can now detect the spatial boundaries of the growth process by finding \emph{spreading speeds} $c$ where $\Re\lambda_*=0$, that is, { 
$\lambda_*=\rmi\omega_*$}. The frequency $\omega_*$ determines the frequency of the invasion process in the leading edge and therefore yields a \emph{linear prediction} for the selected pattern in the wake of the invasion. While there are no mathematical proofs for this selection, there is ample numerical and experimental evidence, as well as a number of criteria which determine when such linear predictions may fail \cite{vS,da,fhs}. 

\paragraph{Double roots and spreading speeds in coupled transport.}
In our example, double roots solve
\begin{align}
d_c(\lambda,\nu)&=(\lambda-c\nu)^2-(\lambda-c\nu)(n_1-n_2)-(n_1+n_2)\nu -\nu^2,\nonumber\\
\partial_\nu d_c(\lambda,\nu)&=-2 c\lambda -(n_1+n_2)+c(n_1-n_2) - 2\nu+2 c^2\nu,\label{e:pdr}
\end{align}
which gives
\begin{align}
\lambda_*&=\frac{1}{2}\left(n_1-n_2 -c(n_1+n_2)\pm 2\sqrt{-n_1 n_2(1-c^2)} \right),		\nonumber\\
\nu_*&= \frac{1}{2(1-c^2)}\left((n_1+n_2)(c^2-1)\mp 2c\sqrt{-n_1n_2 (1-c^2)} \right).
\label{e:pdrsol}
\end{align}
These formulas are valid only for $|c|<1$ since double roots are not pinched for $|c|>1$. Note that $|\Re\nu|\to\infty$ as $c\to 1$, a feature that has received some attention in the context of ``frustrated fronts" in unidirectionally coupled systems; see \cite{dick} and references therein.

We next describe intervals of speeds where $\Re\lambda_*>0$, which in turn are bounded by the linear spreading speed. 

First, suppose that $n_1 n_2<0$, such that the discriminant is positive. Then instability corresponds to $\lambda_*^+>0$, which, after a short calculation is found to be equivalent to $n_1>0>n_2$ (linear instability) and $c\in (-1,1)$. In other words, the spreading occurs with characteristic speed in both directions and is not oscillatory in this case. 

Next, suppose that $n_1n_2>0$, such that the discriminant is negative and the $\lambda_*$ are complex. Then instability  corresponds to 
$n_1-n_2>c(n_1+n_2)$. For $n_1,n_2>0$, this gives $c<\frac{n_1-n_2}{n_1+n_2}$, for $n_1,n_2<0$ the reverse inequality. 

Summarizing, we have instability intervals
\begin{itemize}
\item  $n_1>0$, $n_2>0$: $-1<c<\frac{n_1-n_2}{n_1+n_2}$;
\item  $n_1<0$, $n_2<0$: $\frac{n_1-n_2}{n_1+n_2}<c<1$;
\item  $n_1>0$, $n_2<0$: $-1<c<1$;
\item$n_1<0$, $n_2>0$: $c\in\emptyset$ (stable),
\end{itemize}
where the last set of conditions refers to a regime in which the linear instability decays in a translation-invariant $L^2$-norm, hence in any choice of comoving frame; see the calculation in Fourier space in \eqref{e:dislam}.

From the perspective of wavenumber selection, the instability is oscillatory at the spreading speed only when $n_1n_2>0$, at the ``right edge'' of the spreading region when $n_1,n_2>0$, and at the left edge of the spreading region when $n_1,n_2<0$. At those speeds, $\omega_*=\Im\lambda_*=\sqrt{n_1n_2(1-c^2)}$, which evaluates to 
\begin{equation}
\label{e:omsel}
\omega_*=2\frac{n_1n_2}{n_1+n_2}.
\end{equation}
When $n_1,n_2>0$, these oscillations create periodic wave trains propagating 
to the left at the right edge of the spreading region. Since the speed of 
propagation of the wave trains is the characteristic speed $c=-1$, the wave 
trains can be written in the form $u(t,x)=u_\mathrm{per}(k(x+t))$, with 
$u_\mathrm{per}$ being $2\pi$-periodic in its argument, and  $k$ denoting the 
spatio-temporal wavenumber. In a frame moving with speed $c_*$, $\xi=x-c_*t$, 
we have  $u(t,x) = \tilde{u}_\mathrm{per}(k\xi + k(1+c_*)t)$, with frequency $\omega_*=k(1+c_*)$. Similarly, at the left edge of the spreading region, we find $\omega_*=k(-1+c_*)$. 
Summarizing we find the predicted wavenumber emerging from one side of the spreading region (where spreading speeds are less than one in modulus), after some simplifications, as 
\begin{equation}\label{e:k*}
k_*=n_{{2}} \mbox{ when } n_1,n_2>0,\qquad k_*=-n_1 \mbox{ when } n_1,n_2<0.
\end{equation}
This is a linear prediction for wavenumbers of wave trains emanating in the 
wake of a spreading instability. In the following section, we demonstrate by 
numerical simulations that the linear predictions, both for the absence of wavenumber selection from white-noise perturbations, and for the selected wavenumber for localized or shot-noise perturbations, are accurate. We also demonstrate that the theory for localized perturbations gives good predictions when growth is externally triggered at localized regions of space. 

\section{Wavenumber selection --- scenarios}\label{s:5}
We present results of numerical simulations that illustrate the selection mechanisms described in the previous section. We used first-order upwind finite differences with periodic boundary conditions to discretize in space and an explicit Euler method in time, with grid size $dx=0.01$, step size $dt=0.008$, throughout. We found those to perform better than second- or third-order upwind discretizations, in particular in regard to round-off errors that can cause difficulties when studying growth of perturbations near unstable profiles. We note that the first-order upwind method introduces a small amount of viscosity, which one however could argue is more realistic than dispersion or higher-order viscosity introduced by higher-order upwind methods.  We also compared to Matlab's built-in ODE solvers without significant gain in accuracy. 

In the following, we show results from perturbations of  asymmetric states, from perturbations of symmetric 
states, and  results induced by an externally generated growth of the population 
size at a fixed spatial location.

\paragraph{Invasion of asymmetric states.}

\begin{figure}[h!]
	\centering
	\includegraphics[width=.4\linewidth]{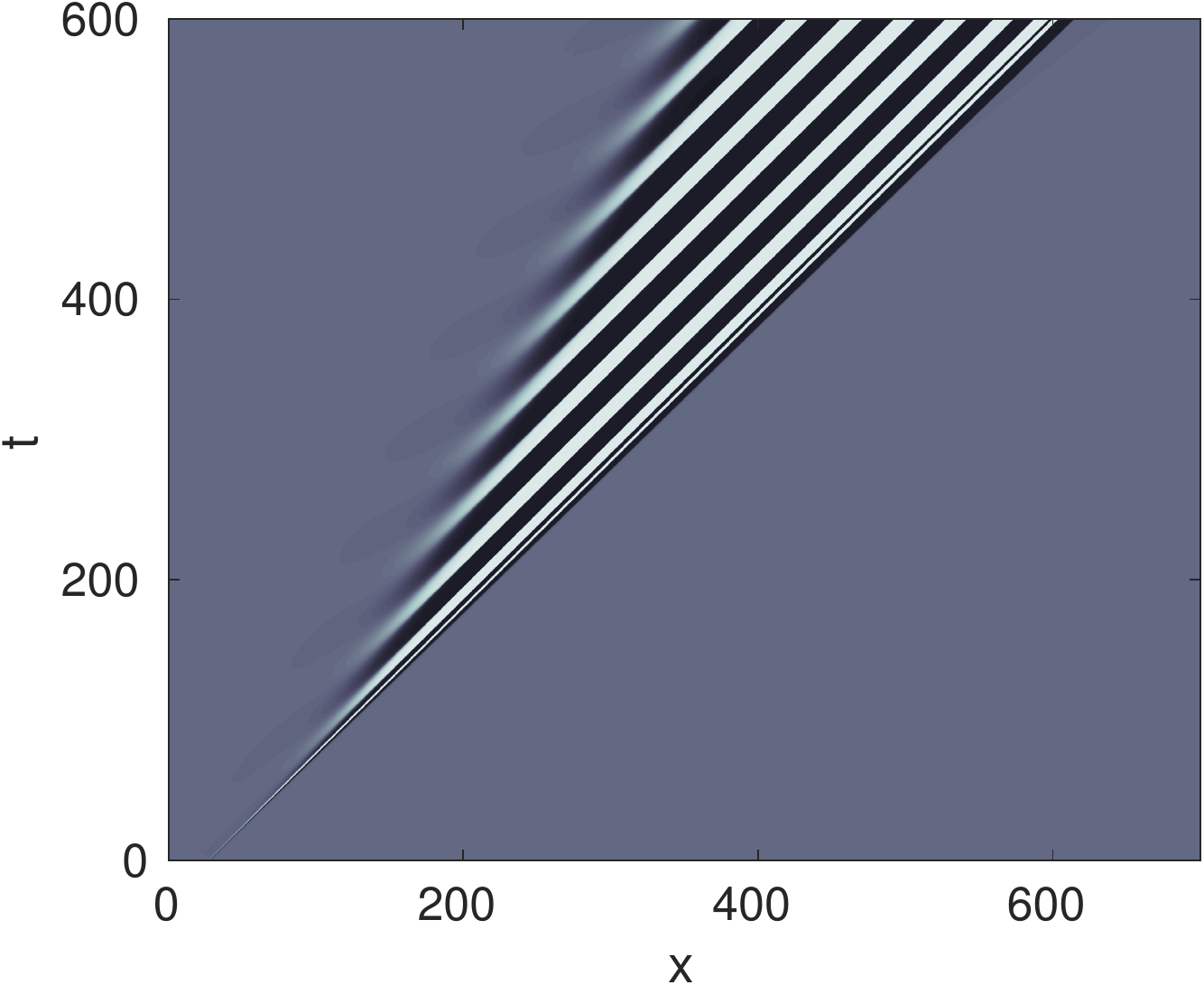}\qquad 
	\includegraphics[width=.4\linewidth]{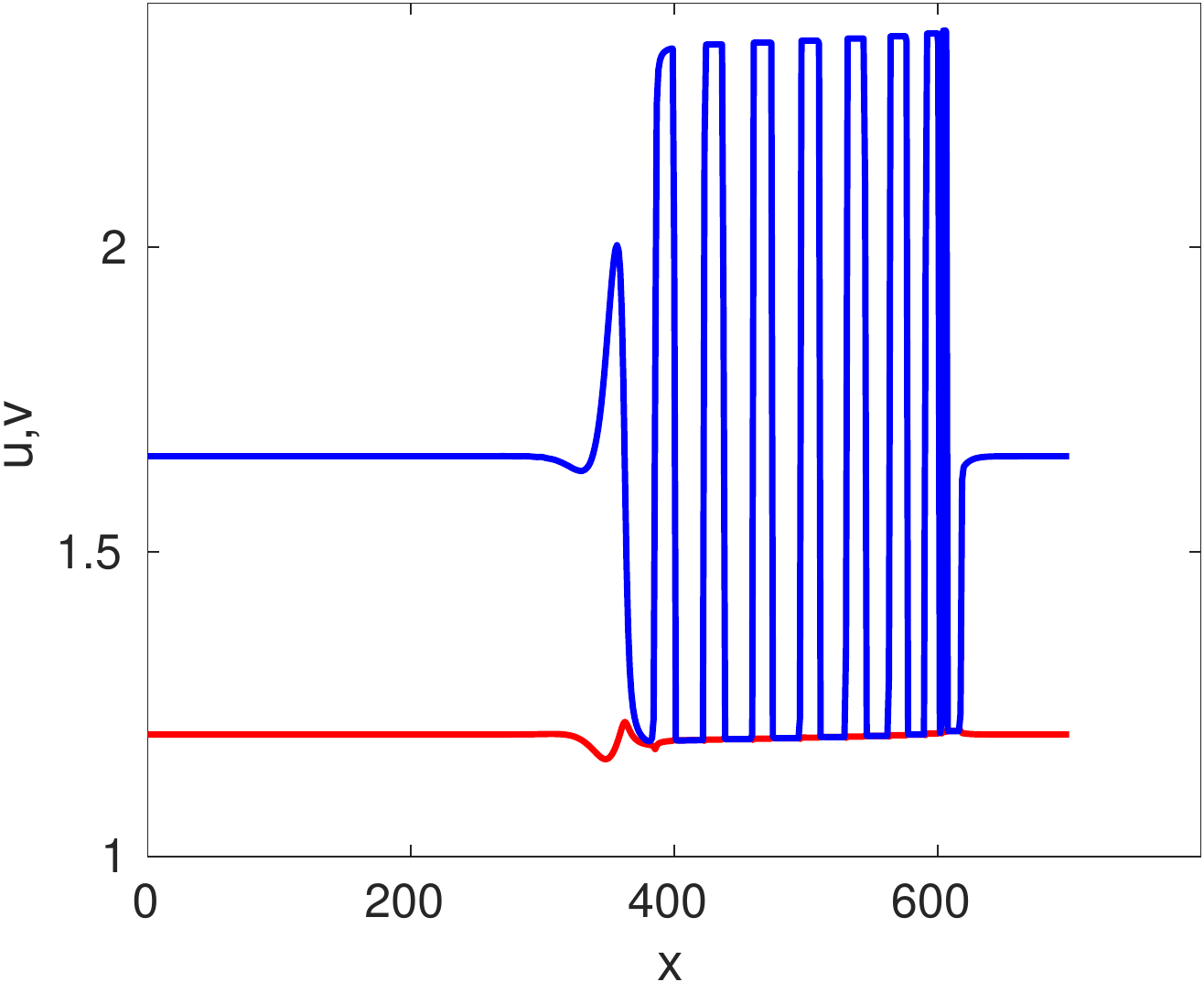} 
		\caption{Space-time plot of the $v$-component and final 
profiles of $u$- and $v$-components after perturbing an unstable asymmetric 
state locally in space. Parameter values are $u_0=1.2,v_0=1.657,\gamma=0.115$ 
({$n_1=-0.1593, n_2=-0.0617$}). The initial perturbation is as in 
\eqref{e:shotshape} with $w=0.3$, $a=1.12,x_0=30$. Note that the 
$u$-component (red) remains almost constant and 
the spreading of perturbation in a cone ($c\in [0.4416,1]$) is as predicted in the previous section.}\label{f:4}
\end{figure}

We first show the effect of a localized perturbation on asymmetric states. 
Our first results in Figure  \ref{f:4} concern $\gamma=0.115$, corresponding to the second diagram of Figure \ref{f:1}.  We apply to the 
constant solution a pointwise zero-mass perturbation in both components,

\begin{equation}\label{e:shotshape}
\delta u(x) =\delta v(x)= \frac{a}{w} \mathrm{sech}\,\left(\frac{x-x_0}{w}\right)\tanh\left(\frac{x-x_0}{w}\right).
\end{equation}
The resulting wavenumber, $k_\mathrm{meas}\sim 0.17$, agrees well with the theoretical prediction $k_*=0.1593$. In particular, we note that convergence of wavenumbers for such invasion processes is expected to be slow, $k(t)-k_*\sim t^{-1}$ \cite{vS}, such that a significant portion of the discrepancy can be attributed to transients. 
We varied the values of $(u,v)$ in the initial condition and found generally satisfactory agreement with the theoretically predicted wavenumbers and speeds.

{For smaller values of $\gamma$, the amplitude of the patterns 
formed in the wake increases. This can be compared to the invariant domain 
condition as it was derived in  
\cite{LutscherStevens}. Here values for $(u,v) \in
[0,s]$ are invariant when  $ (1 + \gamma w^2) (1 + \gamma s^2) \geq sw (1 - \gamma sw)$
for all $w$ with $0 < w < s$, which is true  for any $s$ when $\gamma>1/8$, and for $s<\sqrt{(1-2 \gamma - \sqrt{1-8\gamma})/(2 \gamma (1+\gamma)) } $ for $\gamma<1/8$. Therefore decreasing the value of $\gamma$ leads to more restrictive invariance conditions hinting at the possibility of large amplitude bursts.}
%

\begin{figure}[h!]
\centering
\includegraphics[height=3in]{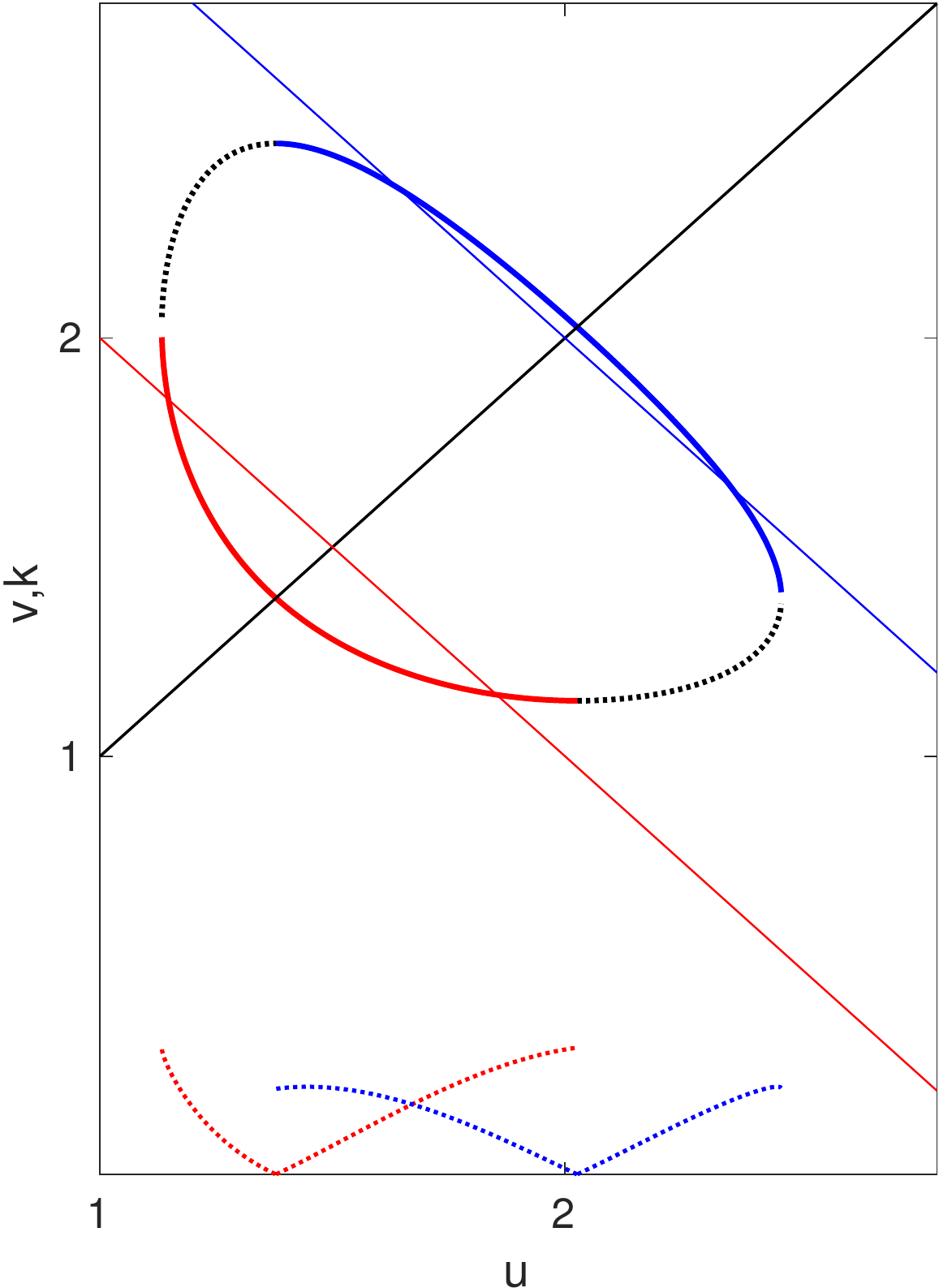}\hfill
\includegraphics[height=3in]{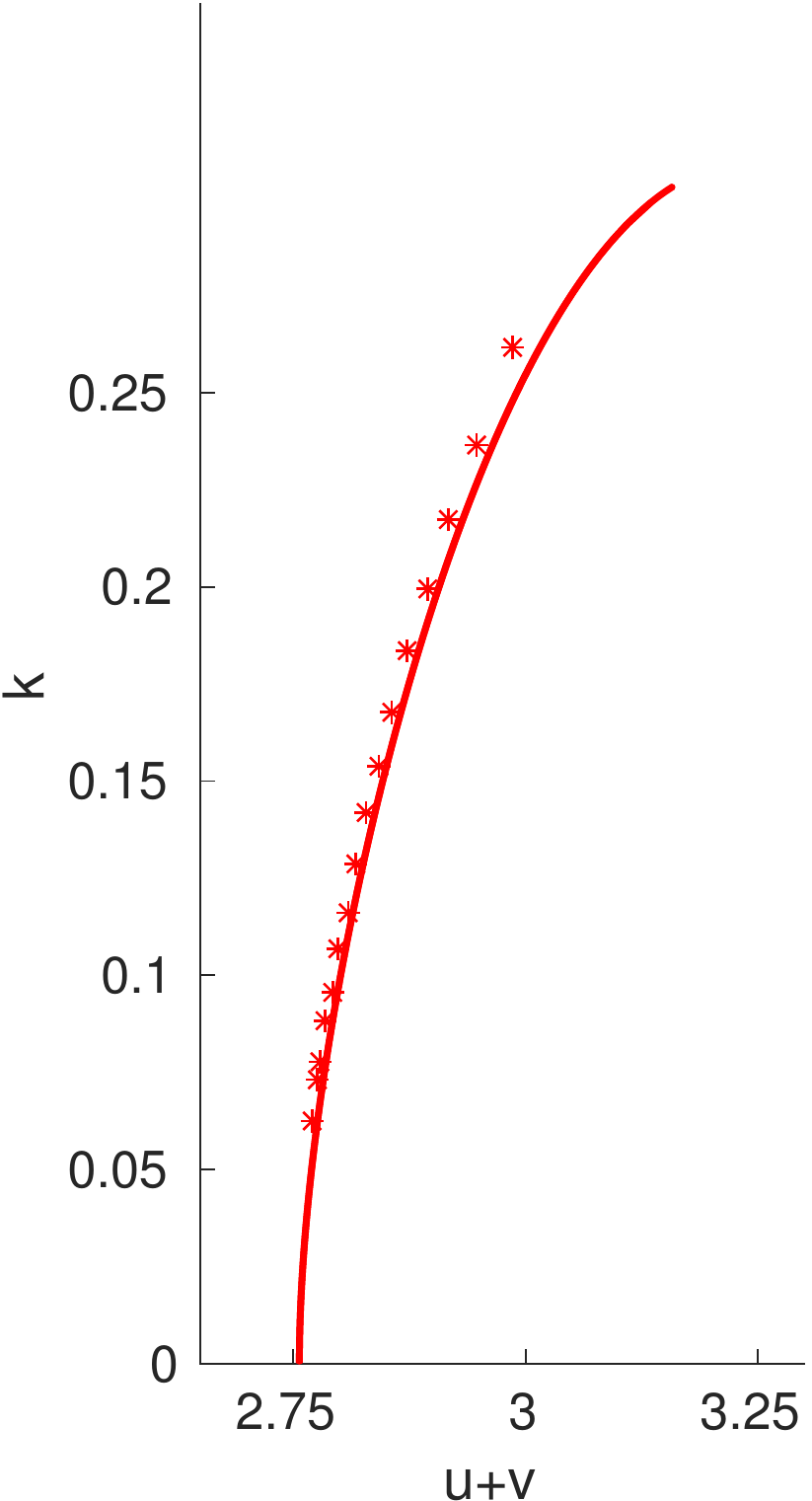} \hfill
\includegraphics[height=3in]{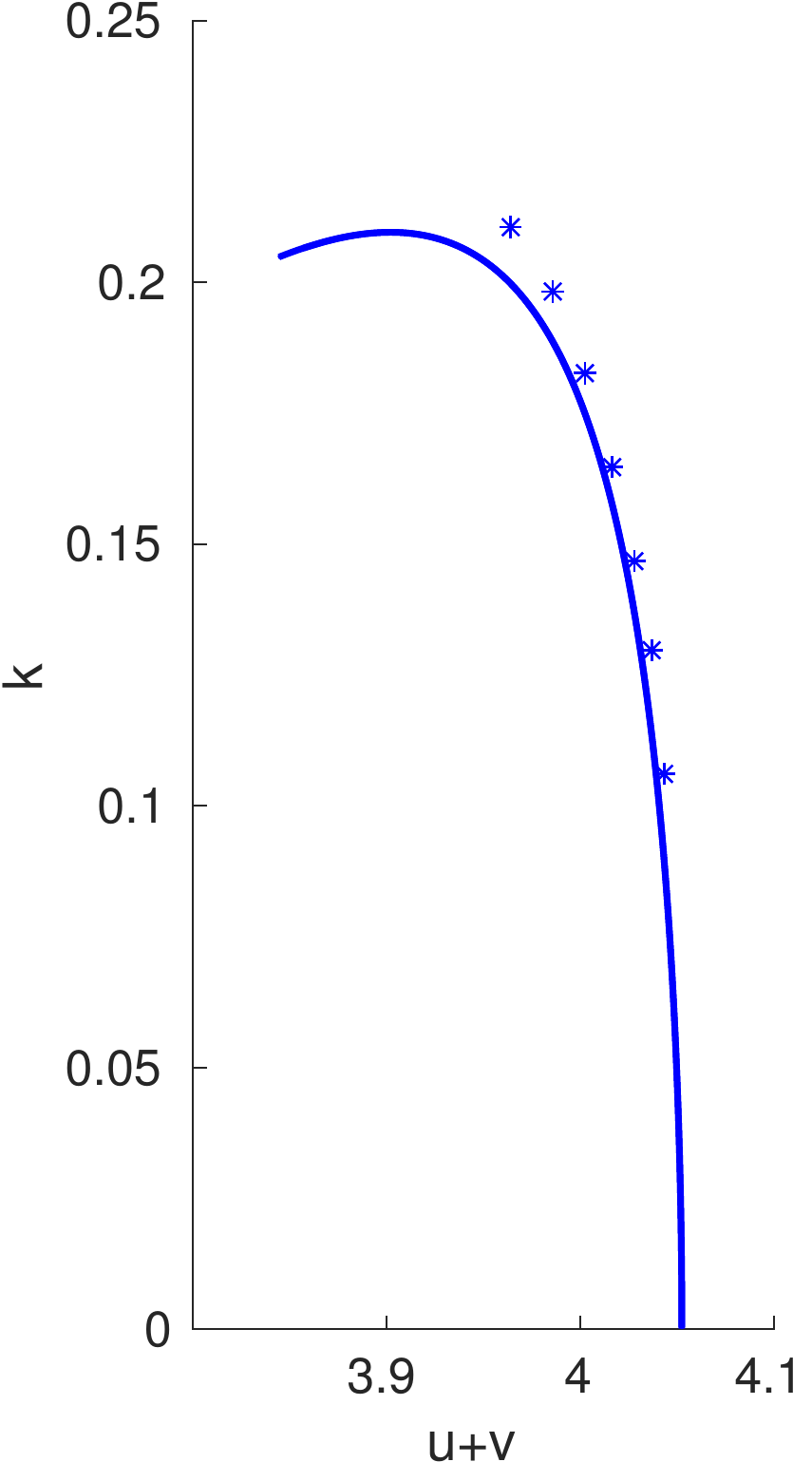}
		\caption{Stable (dashed) and unstable (solid) equilibrium 
branches (red for smaller mass branch, blue for larger mass), and associated selected wavenumbers (bottom curves), illustrating 
the dependence of the wavenumber on the mass (anti-diagonals illustrate 
constant mass). Enlarged plots of selected wavenumbers versus total mass 
(center and right for red (smaller mass) and blue (larger mass) branch), including measured wavenumbers; numerical parameters as 
noted at the beginning of  Section \ref{s:5}, $\gamma=0.115$.} 
\label{f:11}
\end{figure}

\begin{figure}[h!]
	\centering
	\includegraphics[width=.4\linewidth]{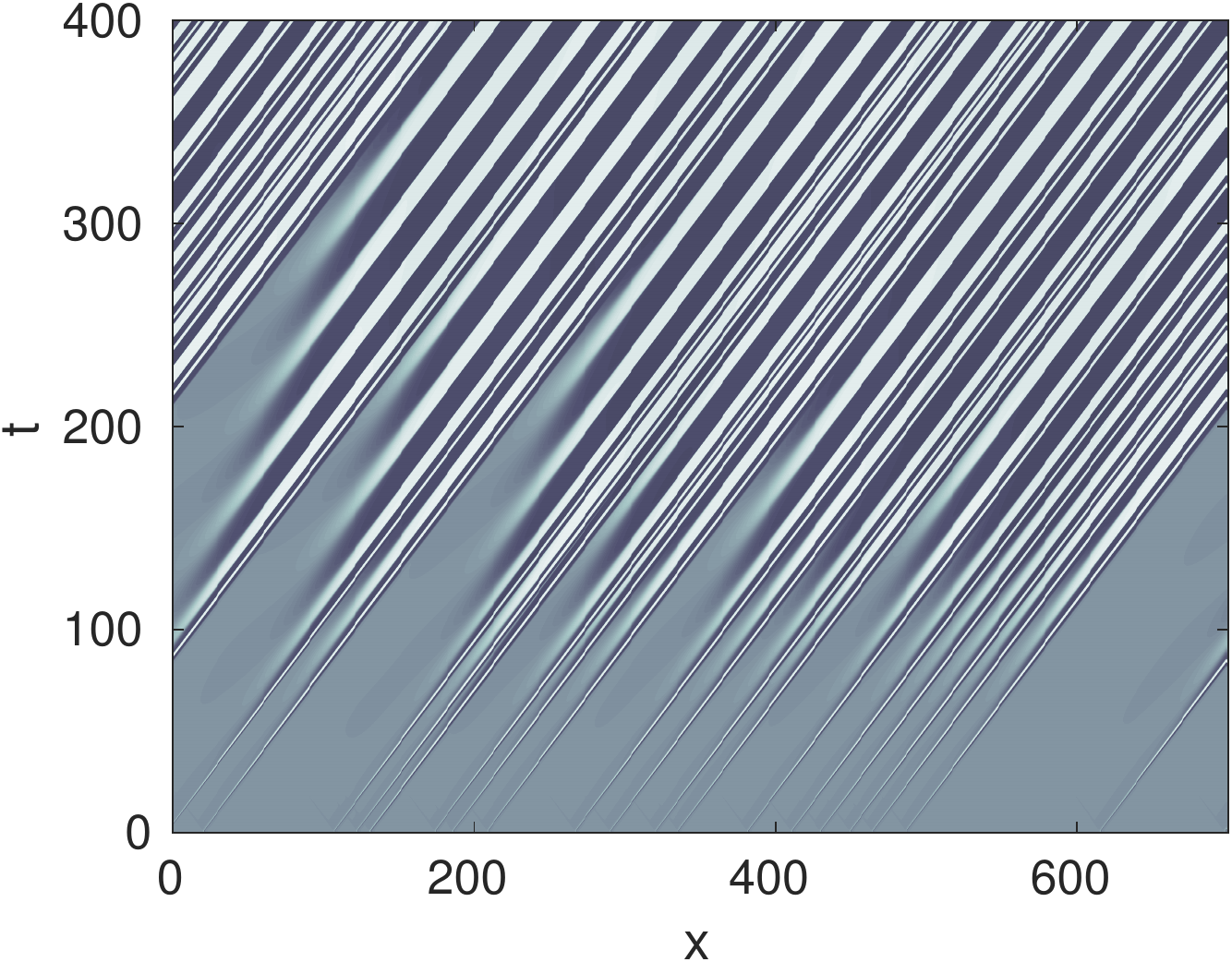}\qquad 
	\includegraphics[width=.4\linewidth]{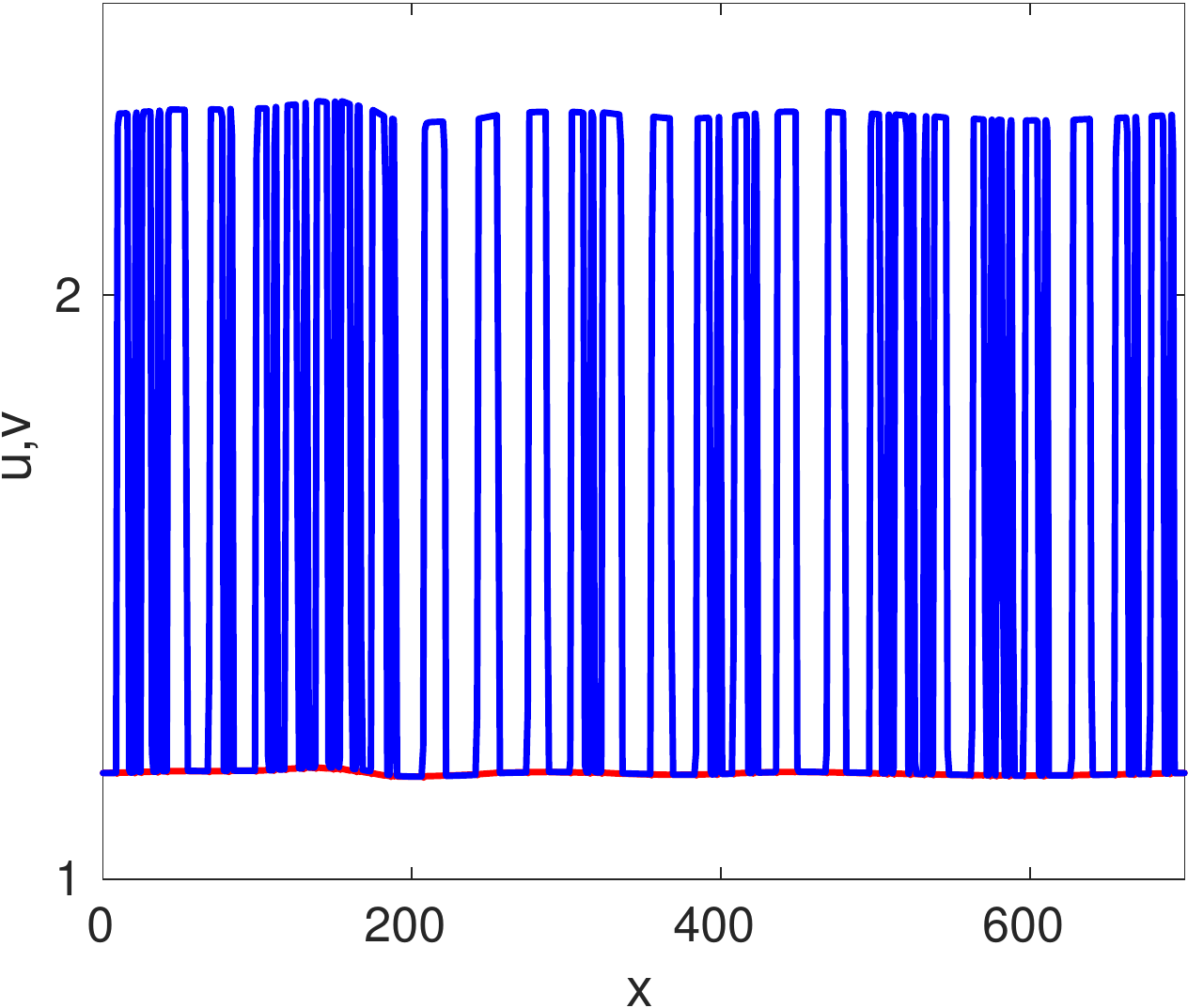} 
		\caption{With parameters as in Figure \ref{f:4}, we look at localized perturbations at 20 random locations. One can distinguish the selected 
wavenumber, as well as irregular parts of the pattern associated with the 
shot locations; space-time contour plot of $v$-compoent (left) and time snap shot $t=400$ (right) of $v$- (blue) and $u$- component (red).}\label{f:6}
\end{figure}

\begin{figure}[h!]
	\centering
	\includegraphics[width=.4\linewidth]{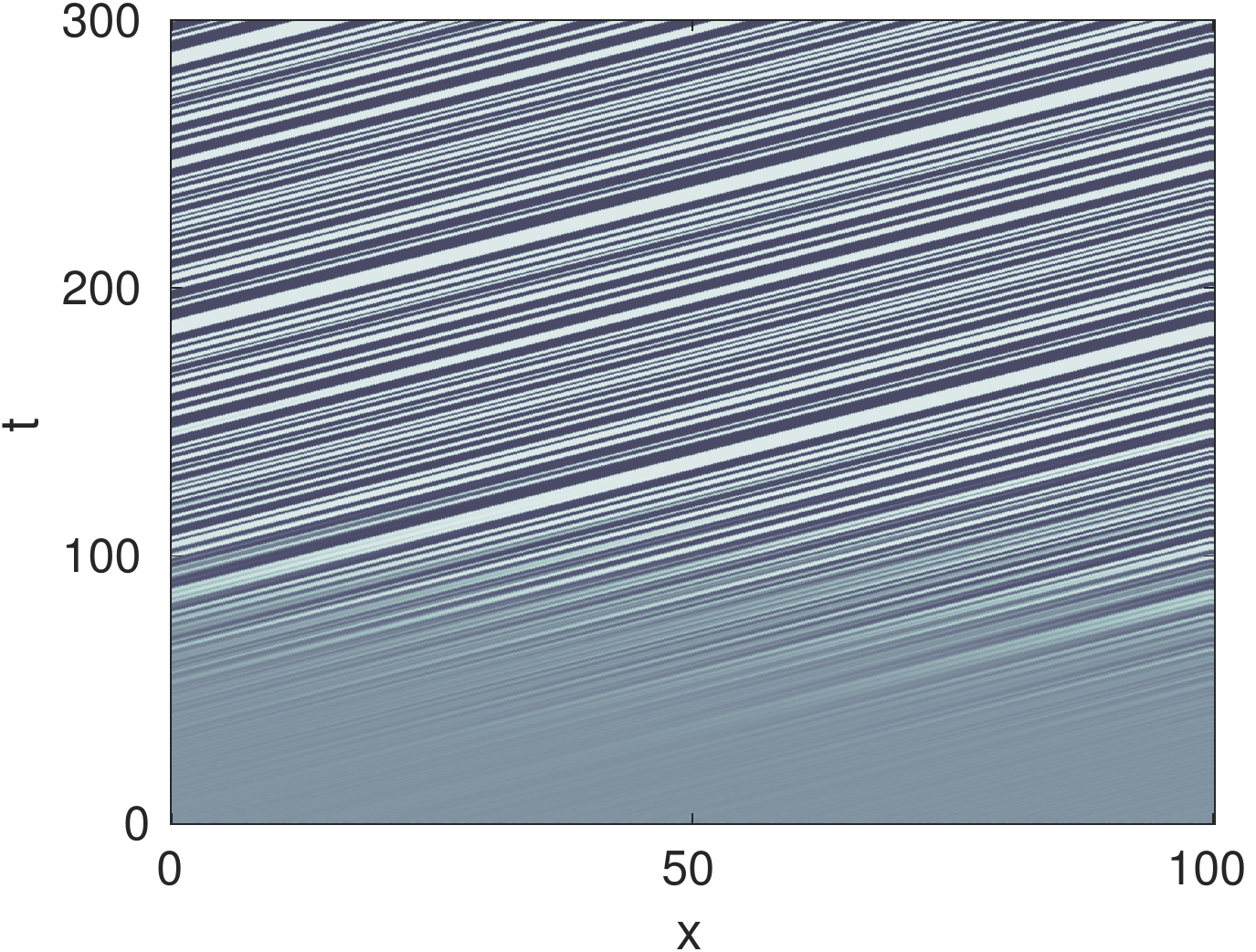}\qquad 
	\includegraphics[width=.4\linewidth]{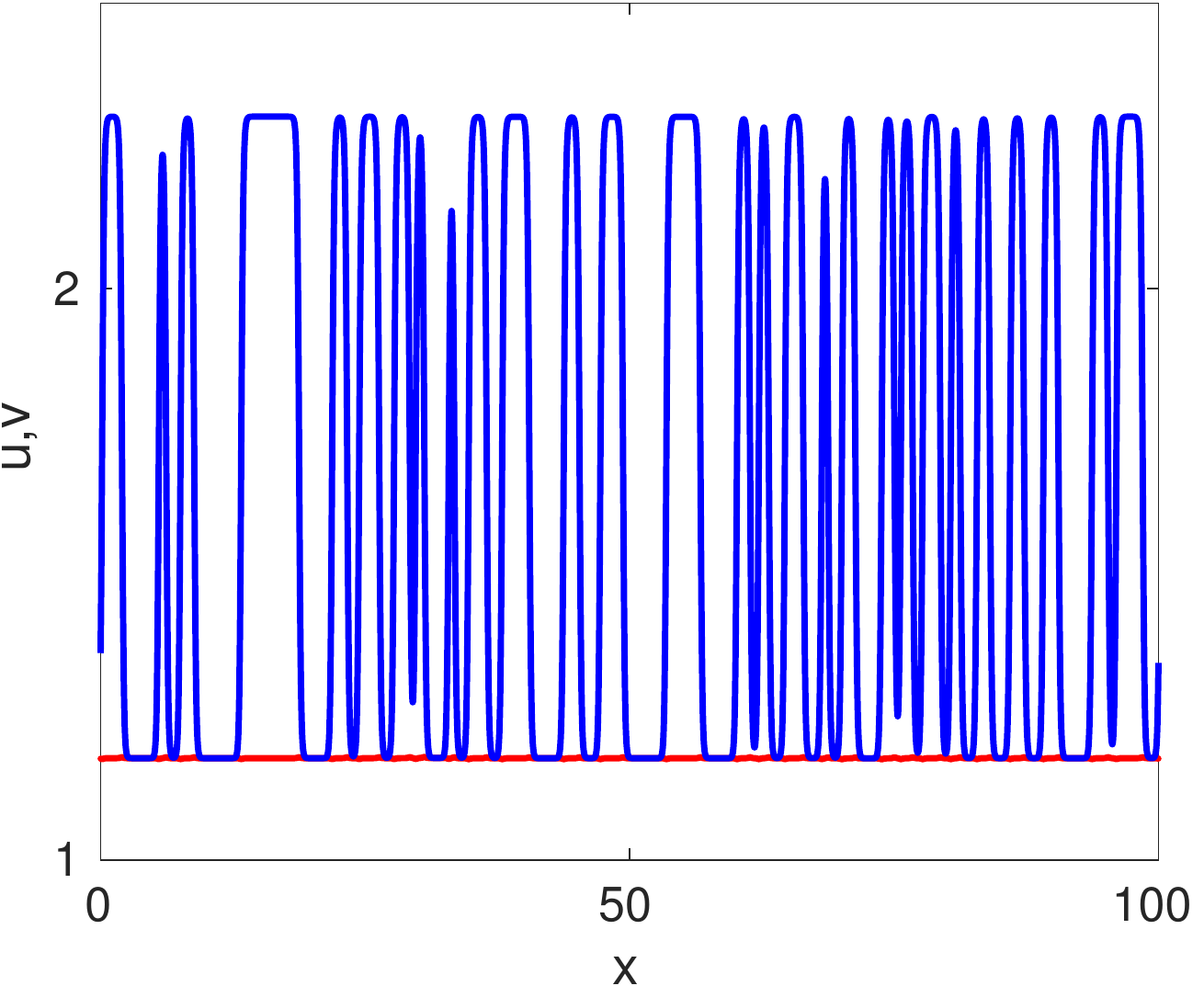} 
		\caption{With parameters as in Figure \ref{f:4}, we start from white-noise perturbations, space-time plot of $v$-component (left), final profiles (right) of $u$ (red) and  $v$ (blue). Wavenumbers in the resulting 
patterns are controlled by numerical diffusion. Note the smaller spatial scale compared to Figure \ref{f:6}.  }\label{f:7}
\end{figure}

In fact, the value of $u$ jumps between the two asymmetric branches associated with a fixed $v$-value, respecting mass conservation. As one can immediately infer from Figure \ref{f:1}, the value of the second asymmetric equilibrium for given $u$ is very large.
\begin{figure}[h!]
	\centering
	\includegraphics[width=.4\linewidth]{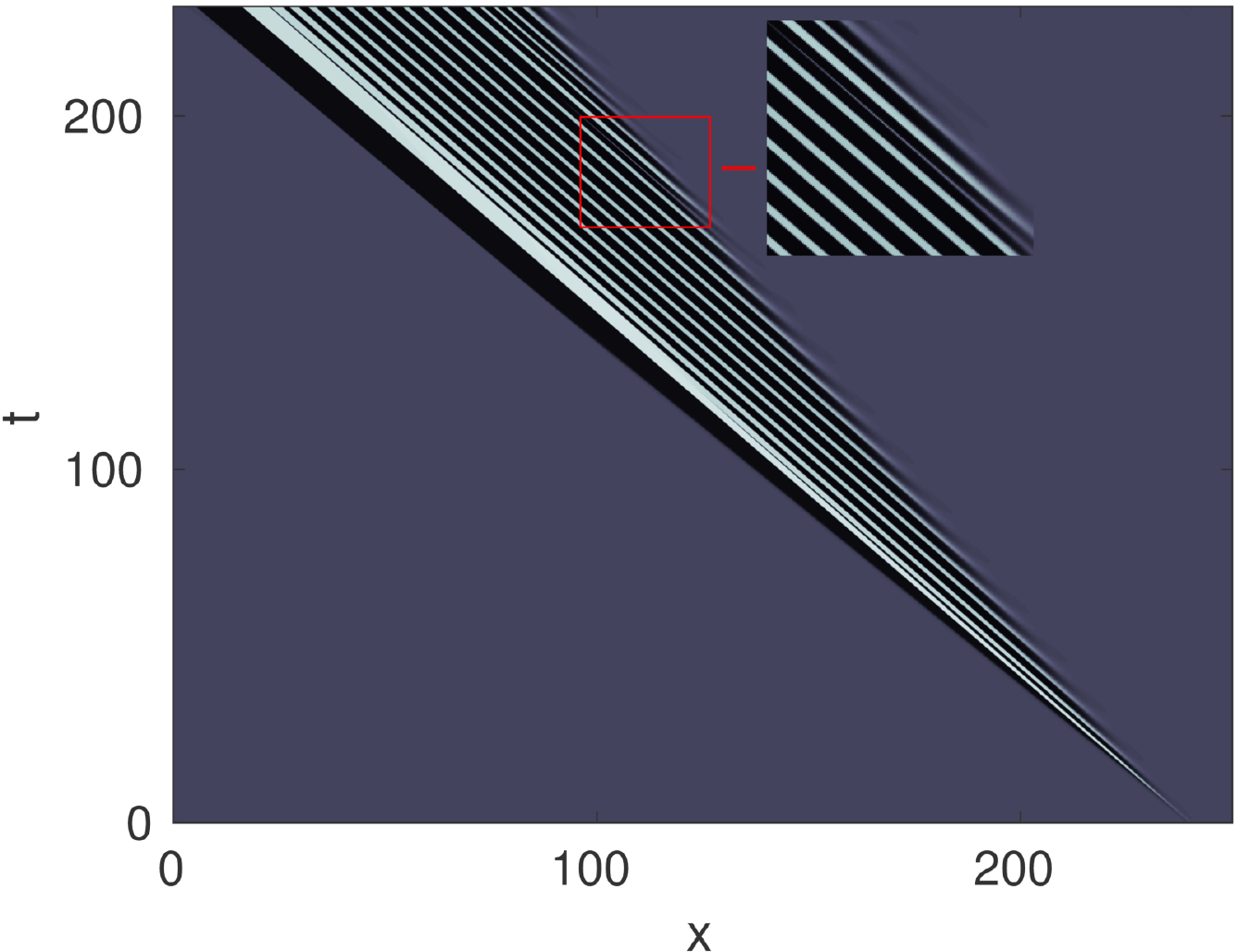}\qquad 
	\includegraphics[width=.4\linewidth]{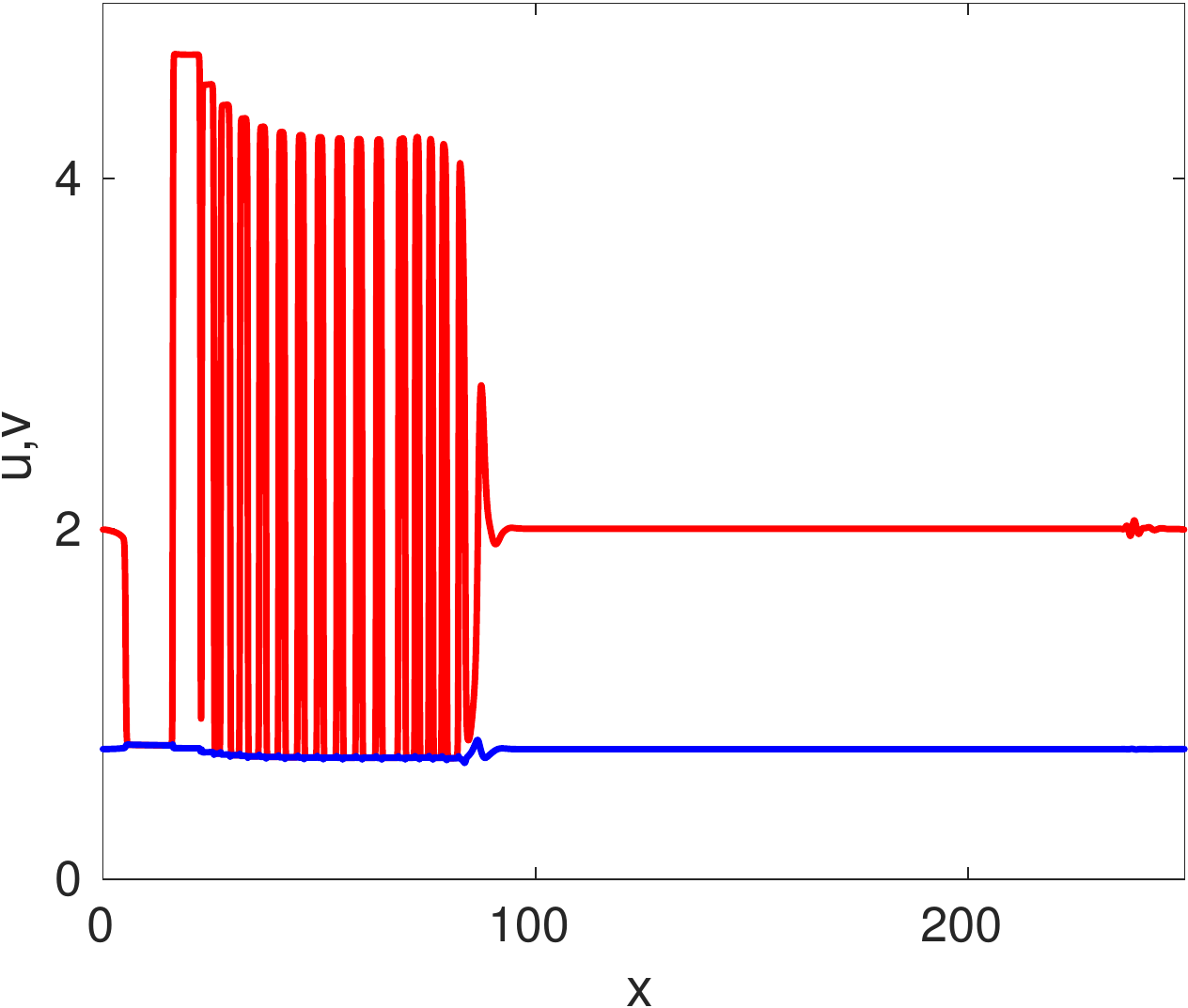} 
		\caption{Space-time plot of the $u$-component and final profiles of $u$- and $v$-components after perturbing an unstable asymmetric state 
locally in space. Parameter values are $u_0=2,v_0=0.741$, $\gamma=0.07$ 
($n_1=0.281, n_2=1.374$), initial perturbation as in \eqref{e:shotshape} with 
$w=0.3$, $a=1.12, x_0=240$. Note that here the $v$-component (blue) remains 
almost constant and the spreading of perturbation in a cone $c\in [-1, -0.660]$
is as predicted in the previous section. {The inset on the left highlights the disappearance of a ripple leading to an effective change in local wavenumber.}}\label{f:5}
\end{figure} 
On the other hand, derivatives $n_j$ are larger and resulting wavenumbers smaller; see Figure \ref{f:5} for an illustration of the results. While generally wavenumbers agree well with the prediction, one can sometimes notice a ``phase slip'' in space-time plots, that is, the nucleation of a ripple
at time $t\sim 150$ which vanishes at time $t\sim 200$; {the inset in Figure \ref{f:5} magnifies this disappearing ripple which is created in the bottom right corner of the inset and vanishes in the top left corner.} We comment on such 
phenomena in the discussion.  A comparison of theoretically 
predicted 
wavenumbers \eqref{e:k*} and wavenumbers observed in direct simulations is shown in 
Figure \ref{f:11}.  The measured wavenumber is typically slightly larger than 
predicted. We attribute this  discrepancy to temporal transients 
since we observed changes in the wavenumber of magnitude comparable to this 
discrepancy across the domain, suggesting that a simulation in 
larger domains over longer time periods would yield more accurate results; see 
for instance \cite{naim} for an example where such comparisons were performed 
to very high accuracy and \cite{vS} for theoretical results on the slow ($1/t$) convergence of wavenumbers. {In practice, this convergence is slow and limited by the accumulation of round-off errors.} For lower masses on the upper branch, we noted failure of nucleation, leading to larger (typically doubled) wavelengths and hence significant discrepancies to the theoretical predictions. 
Perturbations at multiple, random locations (``shot noise'') lead to almost regular patterns, with defects at shot locations; see Figure \ref{f:6}. One 
can observe the expected crossover to irregular patterns when the distance between locations of perturbations approaches the wavenumber. Figure \ref{f:6} also illustrates the absence of a wavenumber selection mechanism for white-noise perturbations. 
Finally, we show simulations with white-noise (random value at grid points) perturbations of amplitude $0.1$. As expected, the scale of patterns is controlled by the (numerical) diffusion; see Figure \ref{f:7}.

\paragraph{Invasion of symmetric states.}

\begin{figure}[h!]
	\centering
	\includegraphics[width=.4\linewidth]{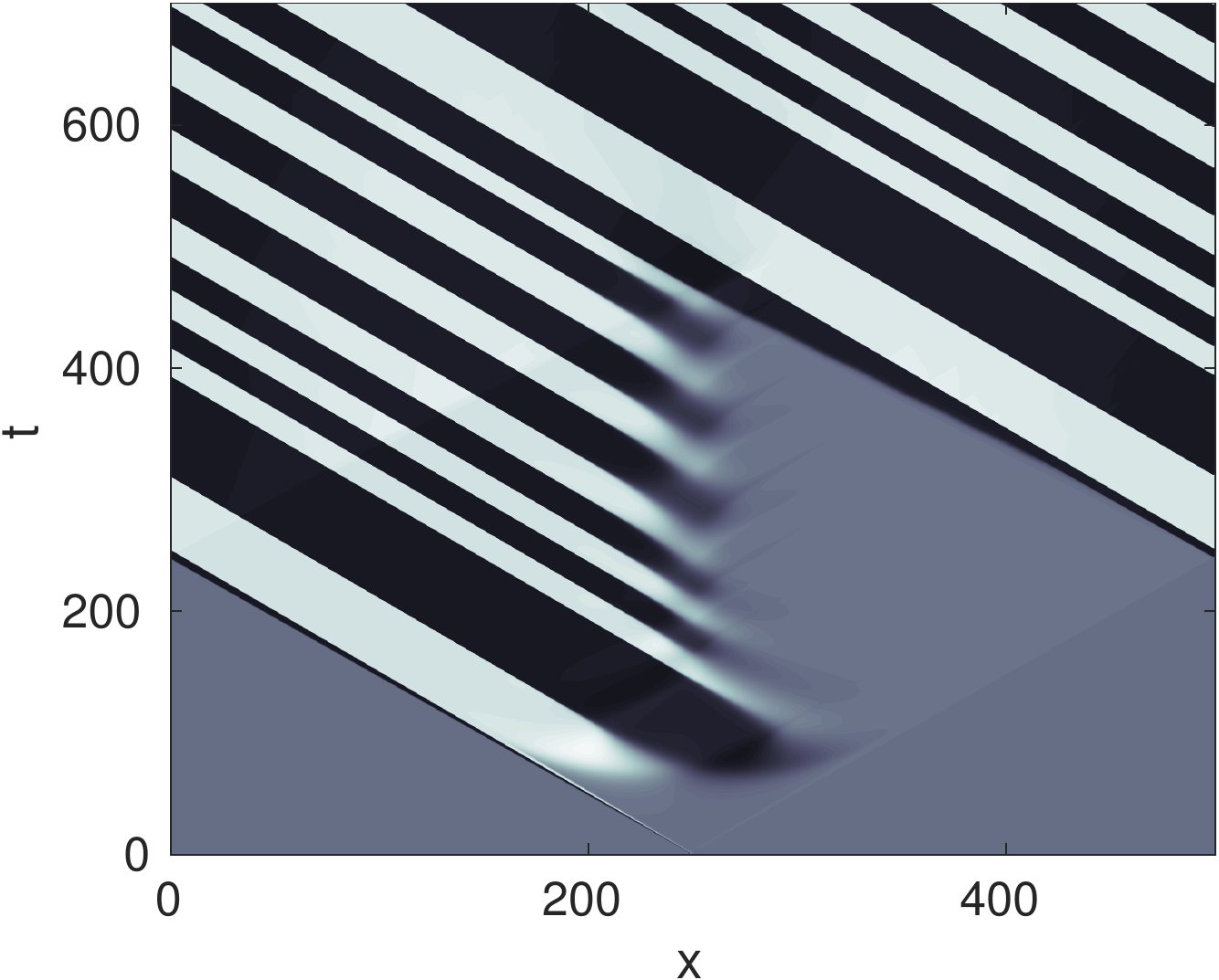}\qquad 
	\includegraphics[width=.4\linewidth]{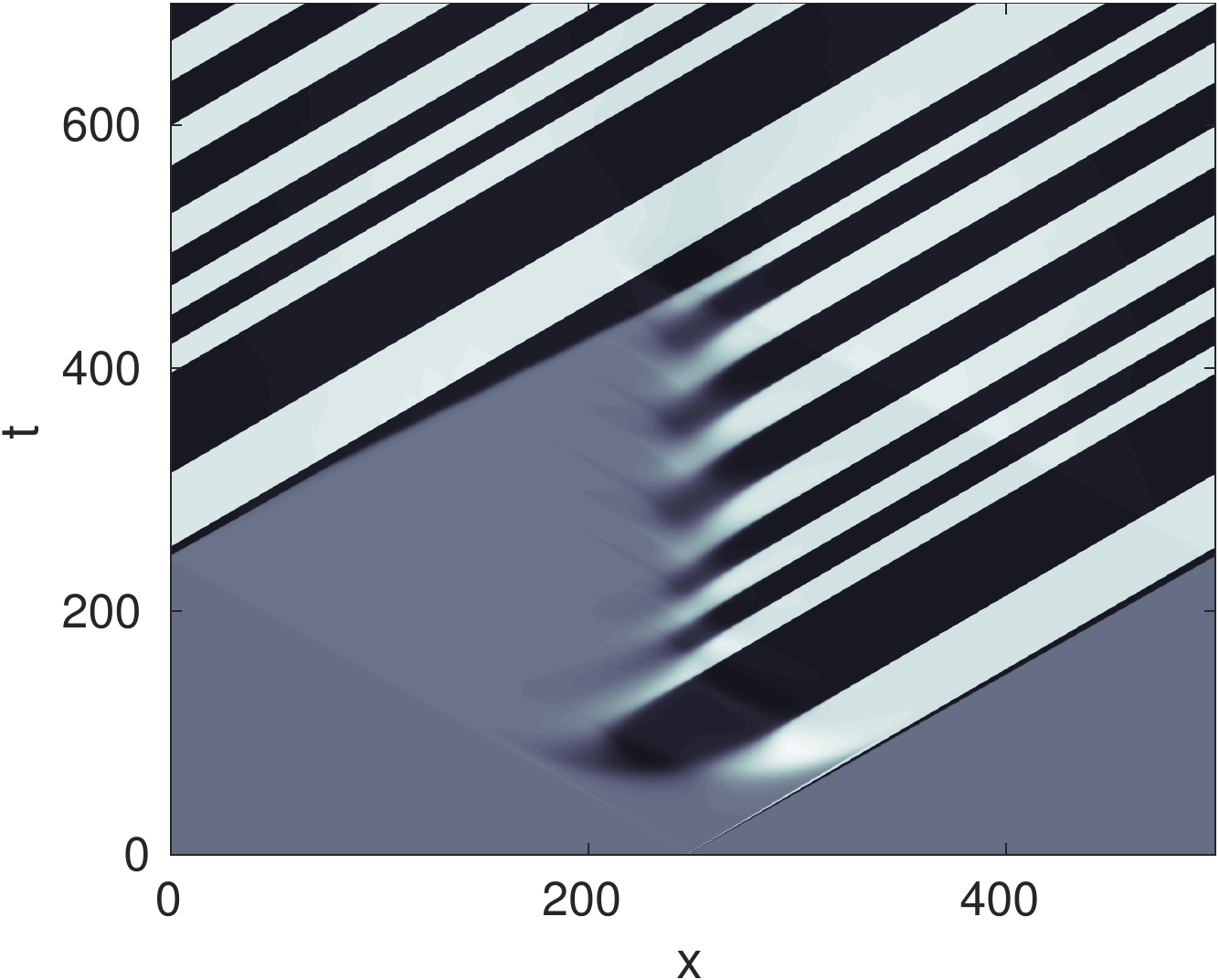} \\
	\includegraphics[width=.4\linewidth]{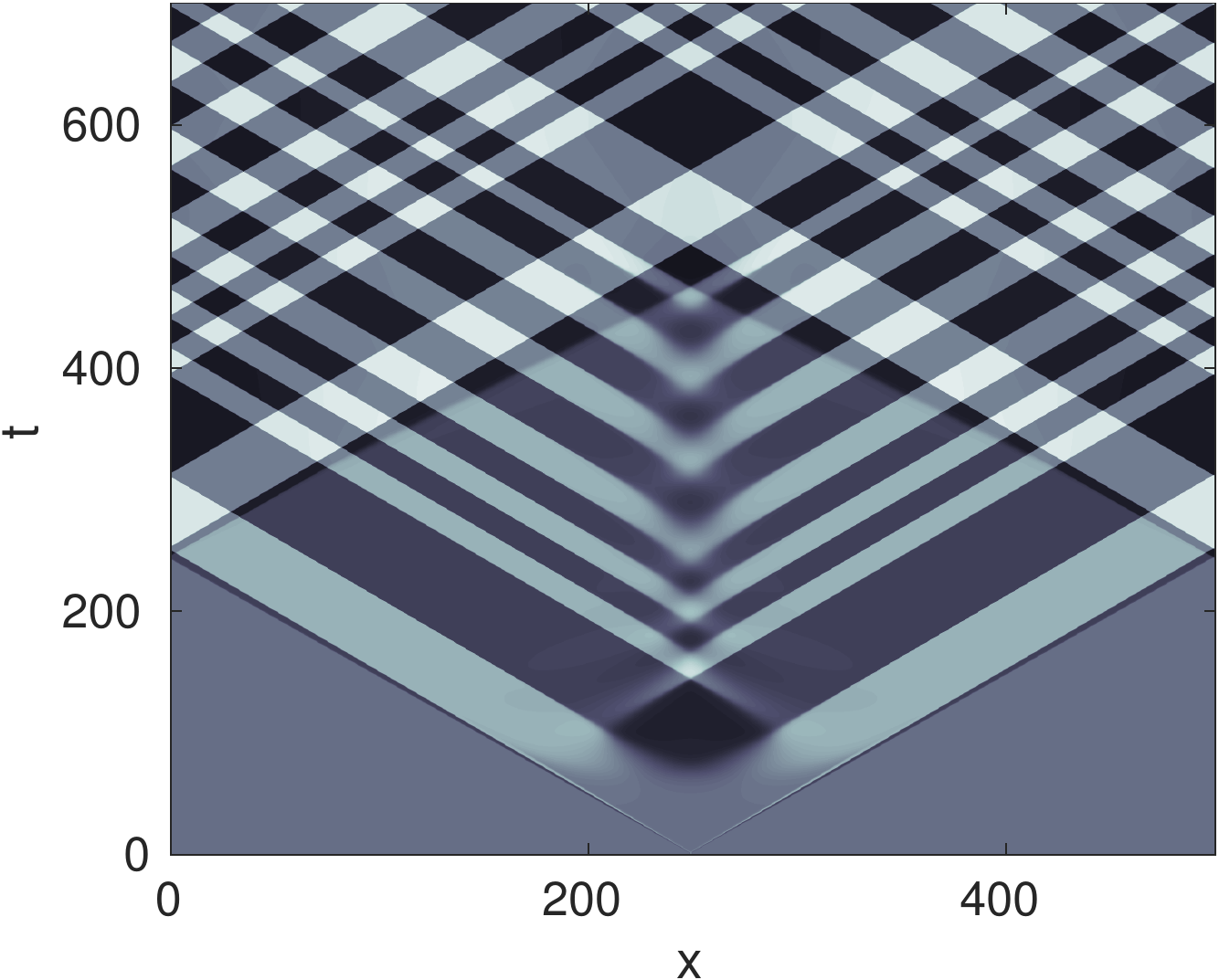}\qquad 
	\includegraphics[width=.4\linewidth]{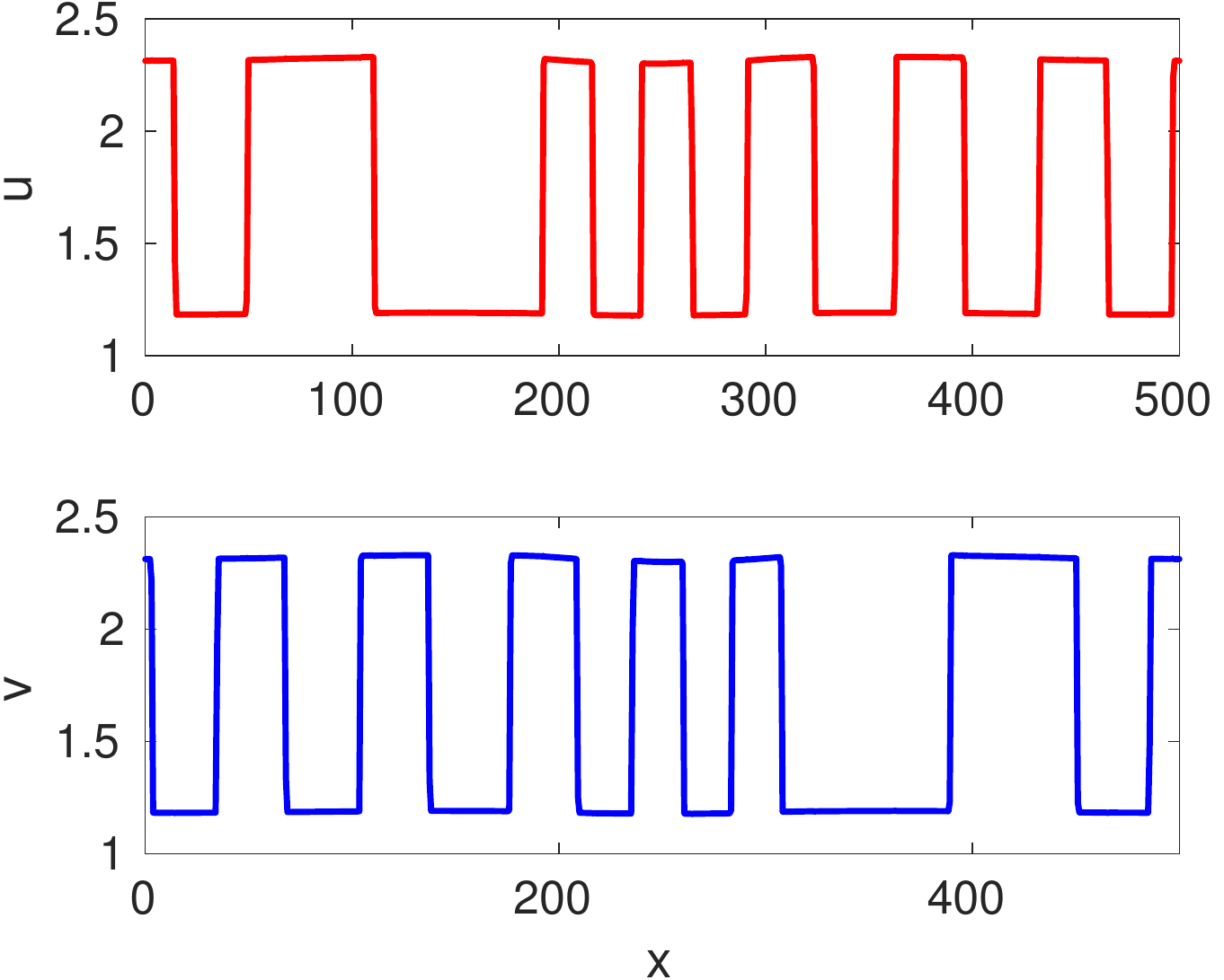} \\
		\caption{Space-time plots of the $u$- and $v$-components (top row); space-time plot of the total concentration $u+v$ and final profiles of $u$- and $v$-components (bottom row). The symmetric state $u=v=1.8$, $\gamma=0.115$ is perturbed by a localized perturbation as in Figure \ref{f:5}.  }\label{f:8}
\end{figure}

{The linear analysis in the Section \ref{s:4} predicts instabilities with zero frequency, that is, spreading of disturbances without oscillations. One indeed observes an instability that simply change the state left in the wake, spreading with the characteristic speed $c=\pm 1$. As discussed in Section \ref{s:3}, such fronts involve a jump in, say, the $u$-component, with $v$ constant across the jump. The resulting state in the wake of the jump is of course asymmetric, and may well be unstable, with small disturbances spreading in an oscillatory fashion. We did observe such a two-stage scenario; see Figure \ref{f:8}. }
{Starting from the initial state  $u=v=1.8$, $\gamma=0.115$, 
we observe a primary change in the $u$-value, say, to $u=1.157$ or $u=2.2362$, which results in 
secondary instabilities with wavenumbers $k=0.2323$ and $k=0.1155$, 
respectively, and predicted wavelengths $L=27$ and $L=54$. The observed 
wavenumber is larger: in fact, the instability is related to the 
instability of the wave train where $u$ is piecewise constant with values 
in $\{1.157,2.2362\}$. It appears to be difficult to derive more accurate predictions for the resulting wavenumbers.} 

We also explored different types of localized perturbations and observed qualitatively similar dynamics. Again, shot noise produces less regular patterns and white noise perturbations of the initial state results in incoherent dynamics without apparent selection of wavelengths.

\paragraph{Growth.} {The selection of wavenumbers induced by the spatial spreading of disturbances is relevant when initial disturbances are well separated in space. A related scenario is when an unstable state is created through a growth mechanism.  We illustrate this here in a prototypical context, again without 
striving for a systematic exploration. We prepare initial conditions as small random (white noise) fluctuations, $(u_0,v_0)\sim 0$.
We then mimic a growth process, where agents $u$ and $v$ are ``created'' 
in the system at spatially localized positions $x=x_j(t)$.} The system with such a spatio-temporal source term reads
\begin{align}
u_t&=+u_x-r(u,v)+r(v,u) + h_u(t,x)\nonumber\\
v_t&=-v_x+r(u,v)-r(v,u) + h_v(t,x).\label{e:ctegr}
\end{align}
We first focus on the simplest case, where $h_j(t,x)=a_j \,  
c_\mathrm{dep}H(x-c_\mathrm{dep}\, t)$, with $\int_\R H(\xi) d\xi =1$, for instance $H(\xi)=(\mathrm{sech}\,(\xi/\delta)/(\pi\delta)$. 

{One would expect that the deposition front leads to the formation of a nonlinear solution $(u_\mathrm{f},v_\mathrm{f})(x-c_\mathrm{dep}t)$, with $(u,v)\to 0$ for $x\to\infty$ and $(u,v)\to (a_u,a_v)$ for $x\to -\infty$. Whenever $(u_a,v_a)$ corresponds to an unstable state, one expects the instability in the wake of this front to result in the creation of spatio-temporal patterns, originating from the front interface $x\sim c_\mathrm{dep} t$.  The initial small fluctuations, superimposed on the nonlinear front profile $(u_\mathrm{f},v_\mathrm{f})$ created by the deposition, help initiate the development of the instability at the front interface. }

\begin{figure}[h!]
	\centering
	\includegraphics[width=.4\linewidth]{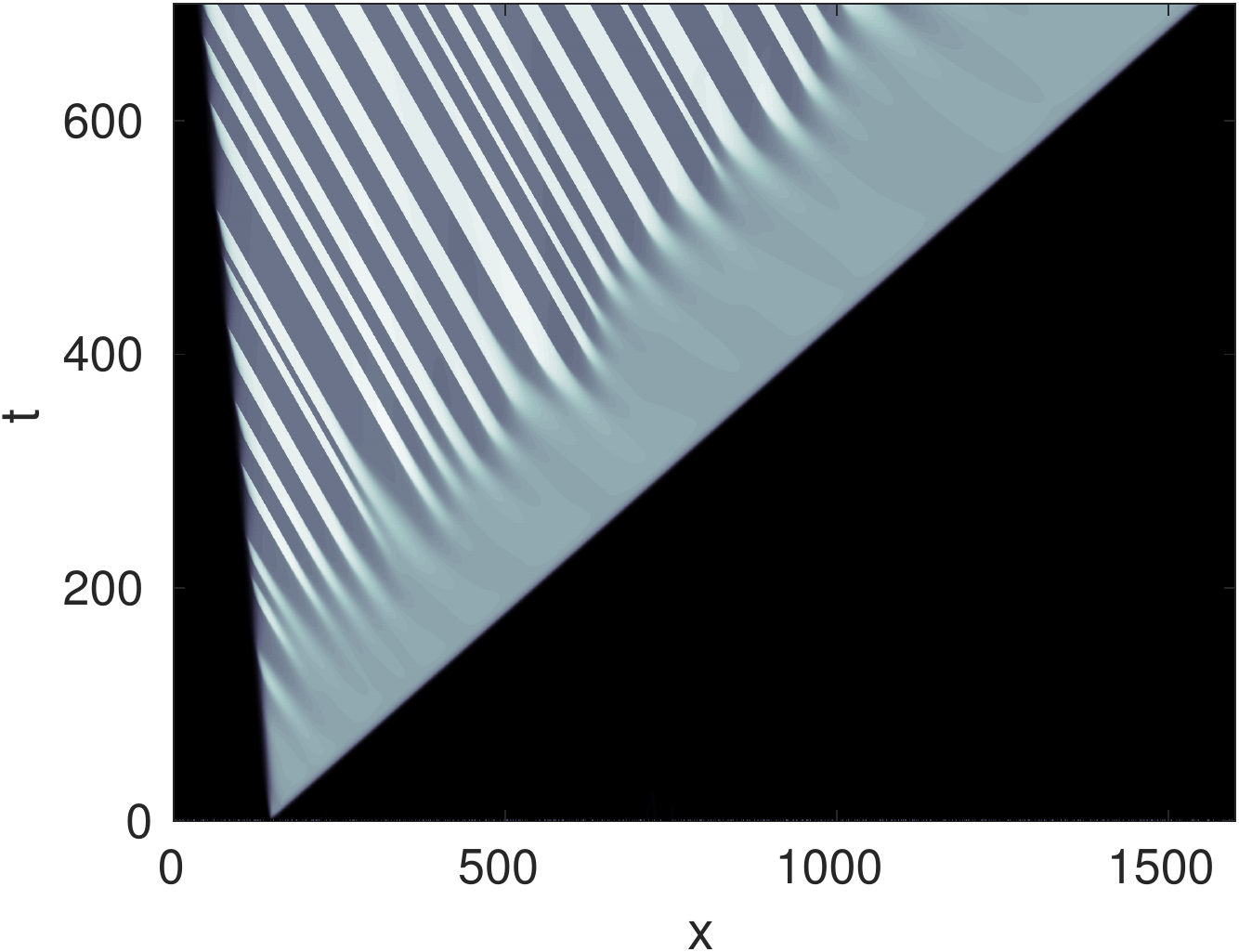}\qquad 
	\includegraphics[width=.4\linewidth]{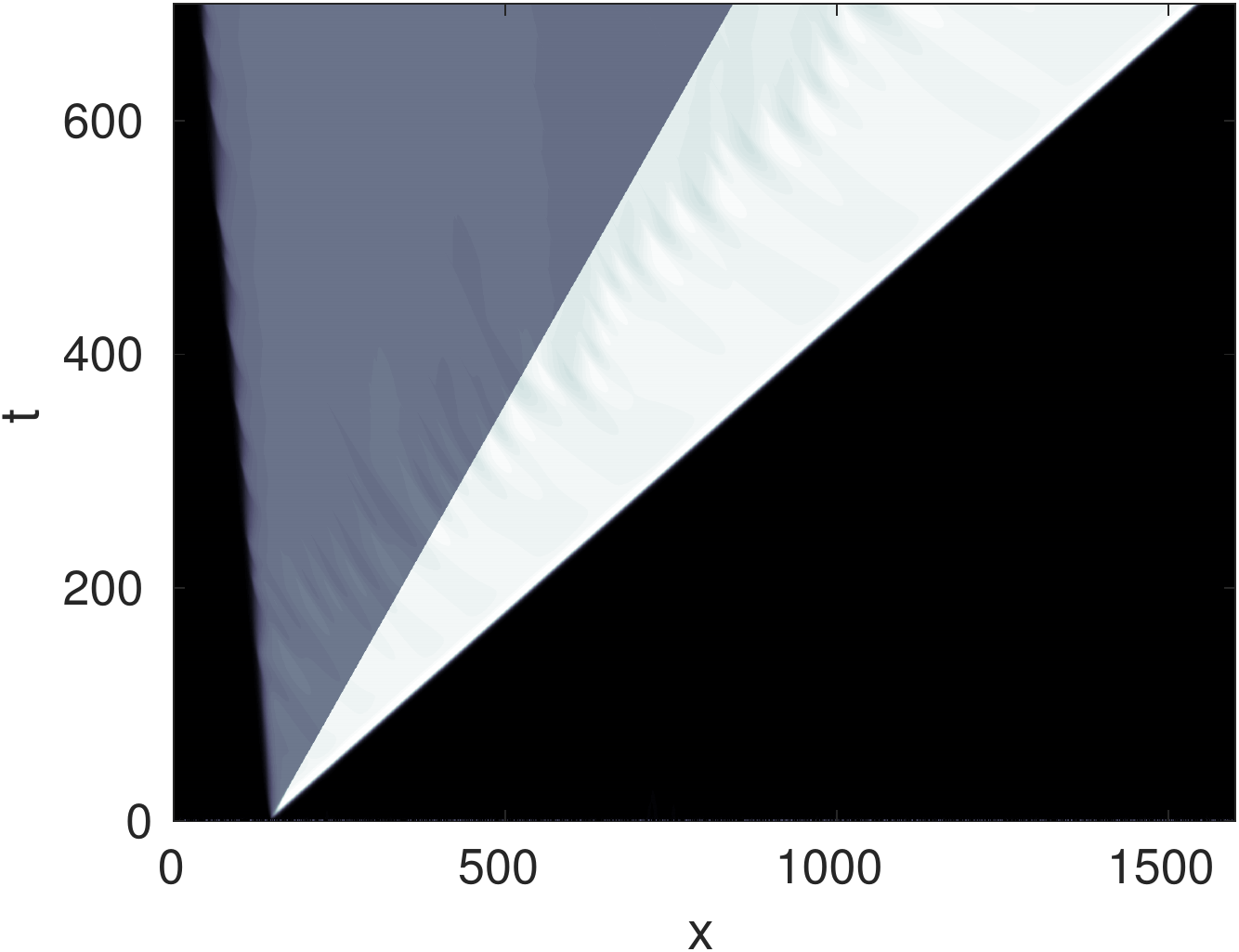} \\
	\includegraphics[width=.4\linewidth]{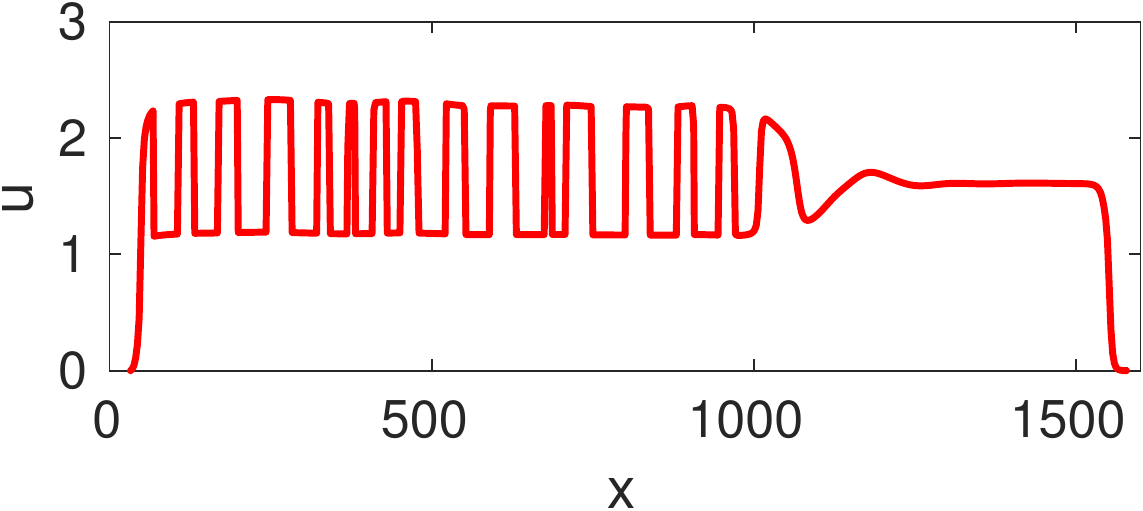}\qquad 
	\includegraphics[width=.4\linewidth]{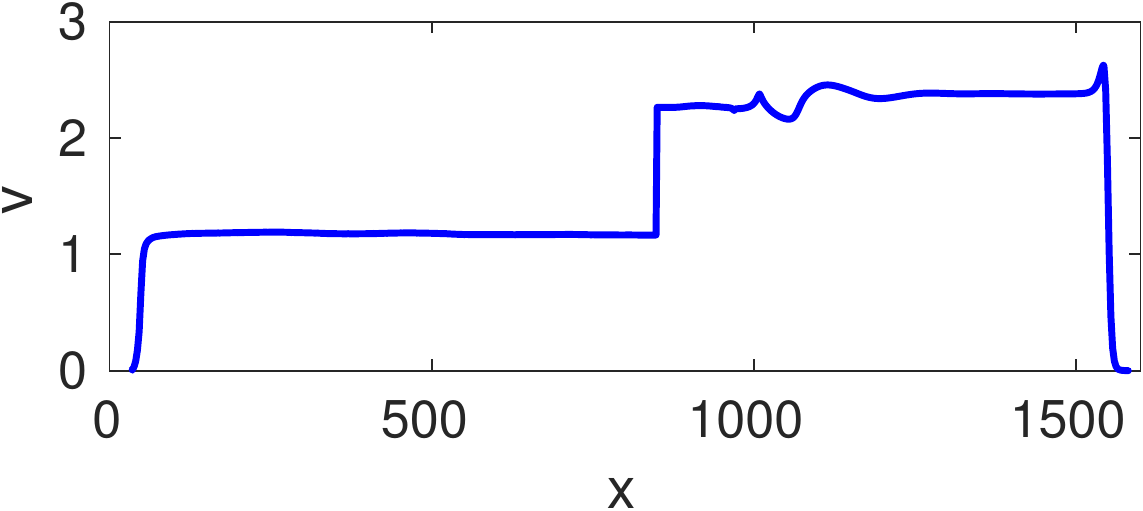}
	
		\caption{Space-time plots of the $u$- and $v$-components for \eqref{e:ctegr} (top row) and final profiles (bottom row) with $\gamma=0.115$, $H(\xi)=\mathrm{sech}\,(\xi/\delta)/(\pi\delta)$, $c_\mathrm{dep}=2$, $a_j=1.8$, $\delta=3$; numerical parameters as described in the text at the beginning of Section \ref{s:5}.}\label{f:9}
\end{figure}

Figure \ref{f:9} shows the result of a numerical simulation. 
Deposition at locations $x=\pm c_\mathrm{dep}t$,  
\begin{equation}\label{e:2source}
h_j(t,x)=a_j^+ c_\mathrm{dep}H(x-c_\mathrm{dep} t)+a_j^- c_\mathrm{dep}H(x+c_\mathrm{dep} t)
\end{equation}
leads to counter-propagating waves in the occupied region. One expects that the selected wavenumber is close to the wavenumber selected by instabilities of the constant state $(a_u,a_v)$ for  speeds $c_\mathrm{dep}\sim 1$, but we did not attempt to derive or validate asymptotics, here; see \cite{gs1,gs2,gs3} for such asymptotics in different contexts. 

\begin{figure}[h!]
	\centering
	\includegraphics[width=.4\linewidth]{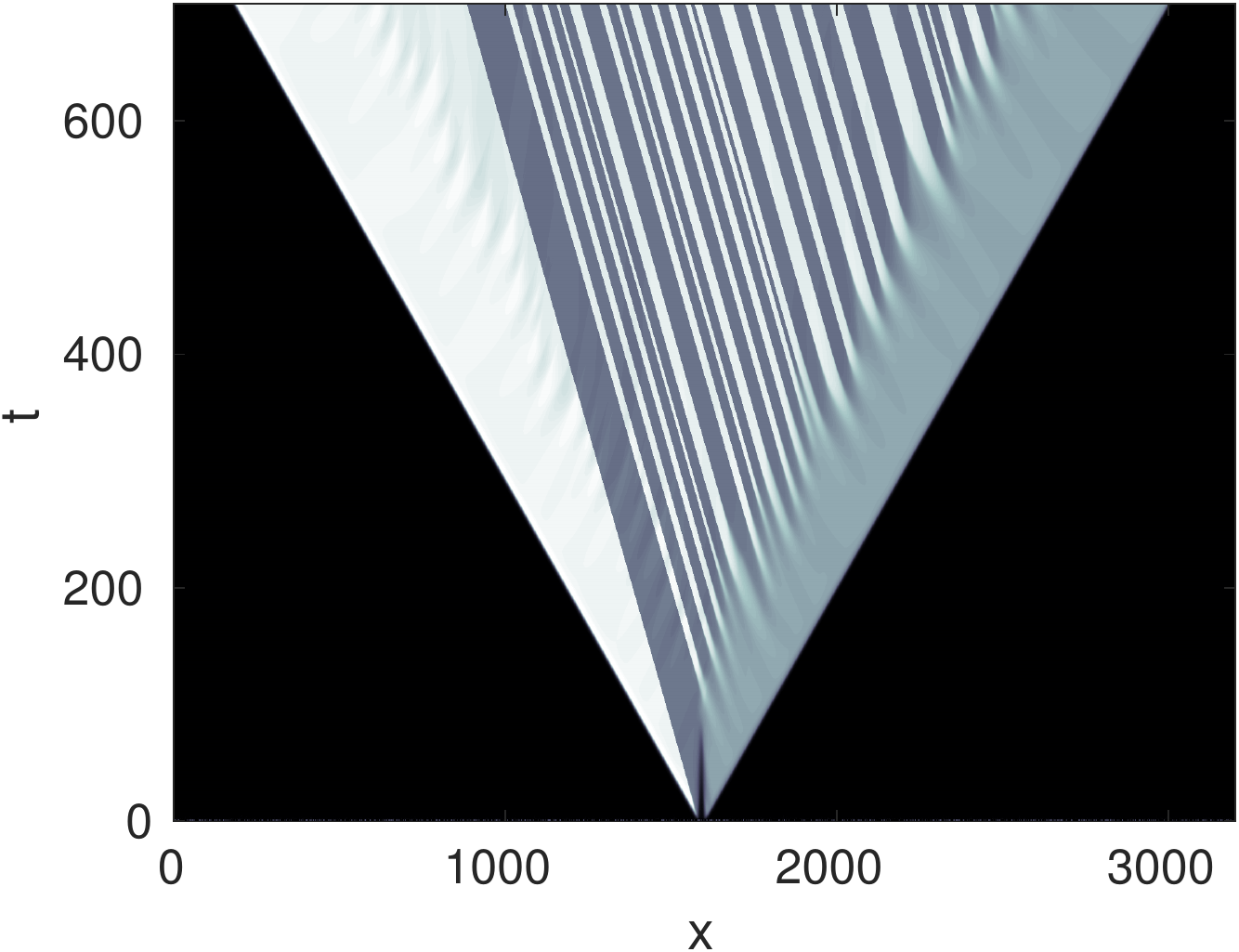}\qquad 
	\includegraphics[width=.4\linewidth]{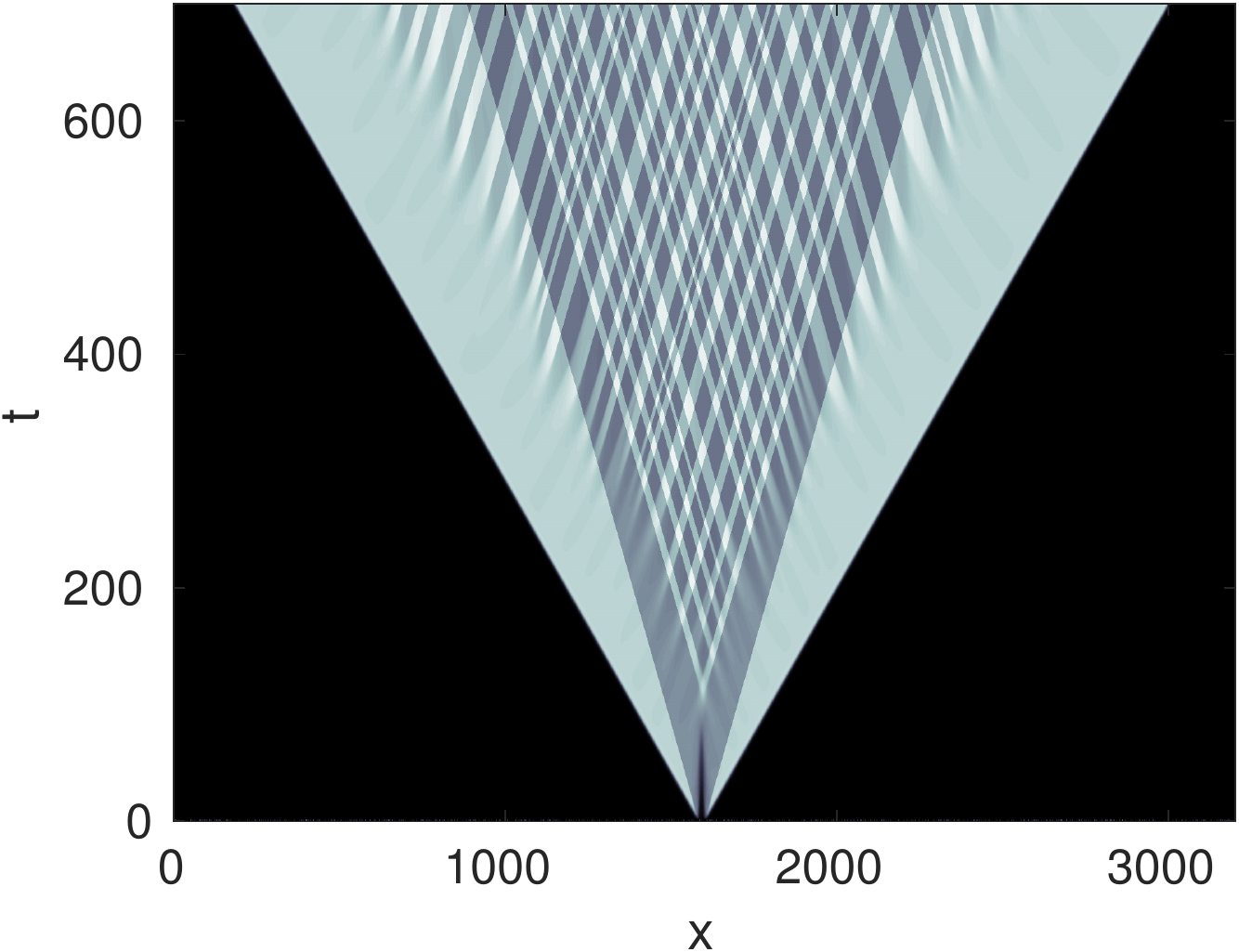} \\
	\includegraphics[width=.4\linewidth]{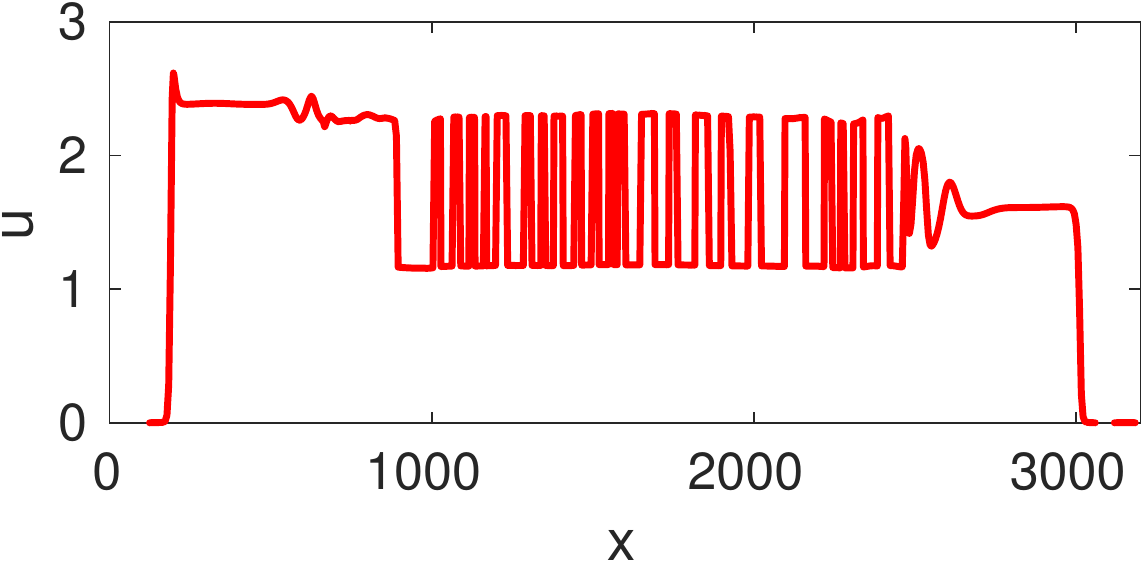}\qquad 
	\includegraphics[width=.4\linewidth]{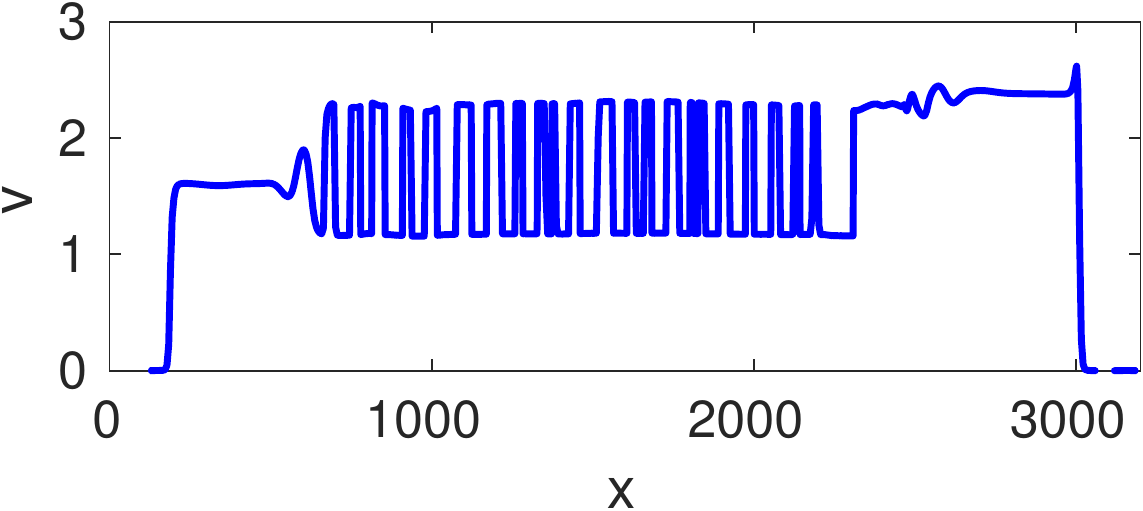} 
		\caption{Space-time plots of the $u$-component, of $u+v$, (top row), and final $u$- and $v$-profiles (bottom row) for \eqref{e:ctegr} with 
$h$ from \eqref{e:2source}, $c_\mathrm{dep}=2$, $a_j=1.8$, 
$\delta=3$; numerical parameters as described in the text at the beginning of Section \ref{s:5}.}\label{f:10}
\end{figure}

Figure \ref{f:10} illustrates the resulting patterns in the symmetric case, $a_j^\pm=a$.
%
%

\section{Discussion}\label{s:6}
We summarize and discuss our results. 

\paragraph{Counter-propagating waves.} We studied a ``minimal'' model for 
run-and-tumble dynamics, with nonlinearity confined to tumbling kinetics. 
Our main results here illustrate a dramatic dependence of the selected pattern on the type of perturbation of the initial state. \emph{White-noise} perturbations of unstable states result in either oscillations of arbitrary fine scales, or long-wavelength modulations of spatially constant states. Localized or \emph{shot-noise} perturbations result in spatially periodic waves with well-defined wavenumber. {While a distinction between wavenumbers selected as the fastest linearly growing mode (relevant for white noise perturbations) and wavenumbers selected through an invasion process (relevant for shot noise perturbations) has been noted early on \cite{dee,vS}, it is often quantitative, changing the wavenumber slightly. }
In that regard, our results are an extreme case, where the nature of the noise enables the selection of wavenumbers. 

On the other hand, there are virtually no systems so far, where a mathematical proof of this wavenumber selection process is available. The best mathematical results rely on the construction of invasion fronts that are periodic in a comoving frame, $U(x-c_* t,\omega_* t)$, and proving stability of those in suitably weighted norms; see for instance \cite{CE,CE2,ES,hs,ga,ch}. Selection of invasion 
fronts from compactly supported initial data has not been shown in pattern-forming systems. It would be interesting to attempt such a study in the present case, that is, study the existence of invasion fronts with speeds and frequency 
from the linear calculus given in Section \ref{s:4}. 

In a different direction, the intriguingly simple structure of our systems insinuates that a more direct, hands-on description might be possible. As an 
example, considering the equation for time-periodic invasion fronts $(u,v)(x-ct,\omega t)$, leads to an evolution equation with ``time'' variable $\xi=x-ct$ and ``space'' $\tau=\omega t$, which is again a coupled transport equation, now with tumbling kinetics $u_\xi=v_\xi$. Also this equation obviously possesses a 
large group of transformations that preserve its structure, $u=\phi(\tilde{u})$, $v=\phi(\tilde{v})$, with $\phi$ being invertible. It seems however difficult to reduce the 
system to a form that is accessible to an explicit analysis. 

Possibly most intriguingly, one would suspect that the \emph{linear} coupled transport equation 
\begin{align*}
u_t=&+u_x+n_1 u + n_2 v\\
v_t=&-v_x-n_1 u - n_2 v,
\end{align*}
should be amenable to a more direct (that is, not reliant on Fourier transform) analysis. In particular, one would hope to derive the selected wavenumber $n_1$ (or $n_2$) in a straightforward fashion.

\paragraph{Experimental implication and validations.} We commented in the 
introduction on experiments showing rippling patterns with a distinct 
wavenumber in myxobacterial colonies. While our model is clearly too simple to 
capture many relevant features, let us briefly discuss 
whether it can capture 
an essential feature of the selection mechanism. A semi-quantitative test case 
is experiments where mutants are introduced into the system that do 
not release the C-signal. Therefore their presence does not induce tumbling 
of opposite moving bacteria,  whereas the mutants themselves 
do turn upon interaction with counter-migrating wildtype  cells. 
The overall population is always kept constant.
Experiments show 
that the higher the density of mutants is in the mixed mutant-wildtype
populations, the larger the wavelength of ripples becomes, until the
pattern disappears completely when about $90\%$ of bacteria in the population 
are
C-signaling mutants, \cite{SagerKaiser94}.

The resulting system for unaltered $(u,v)$ and mutant $(\tilde{u},\tilde{v})$ bacteria is 
\begin{align*}
u_t&=+u_x-ug(v)+vg(u),\\
v_t&=-v_x+ug(v)-vg(u),\\
\tilde{u}_t&=+\tilde{u}_x-\tilde{u}g(v)+\tilde{v}g(u),\\
\tilde{v}_t&=-\tilde{v}_x+\tilde{u}g(v)-\tilde{v}g(u),
\end{align*}
which is in skew-product form, that is, the $(u,v)$ dynamics do not depend on the $(\tilde{u},\tilde{v})$-dynamics. In particular, selected wavenumbers depend on the selection through the non-mutant population. Fixing the total population in the system, one can therefore view the introduction of mutants as equivalent to a reduction of total mass in a pure non-mutant population. 
This dilution experiment was considered as test case for the rippling
models in \cite{LutscherStevens, AlberJK, SNZO06, PrimiStevensVelazquez}

The dependence of wavenumbers on the equilibria and the total mass was shown in Figure \ref{f:11}. One notices that wavenumbers are decreasing with mass on the branch with lower mass and are (mostly) monotonically decreasing on the branch with higher mass. Pictures for smaller values of $\gamma$ are somewhat similar but more intricate.
%


In summary, these findings agree with the experimentally observed increase 
in wavelength as the percentage of mutants in the population is increased
as given in Figure $7E$ in \cite{SagerKaiser94},
but only within a well defined range of total mass, that is, for masses corresponding to asymmetric equilibria on the lower non-trivial branch. In this respect, the anomalous decrease of the wavelength with increasing mass is associated with the restabilization and subsequent new destabilization of the asymmetric branch. Of course, tuning $\gamma$ or possibly altering kinetics further, one can arrange for the lower asymmetric branch to exist for a large range of total masses, and simulataneously push the upper unstable asymmetric branch to very 
large, potentially unrealistic masses.

\paragraph{Modulation.} 

In the marginally stable regime, $n_1<0$, $n_2>0$,
the dispersion relation suggests exponential damping of modes with $k\neq
0$, such that one {would like to expand in terms of long-wavelength modes.
One therefore expands the dispersion relation} to find near, say, $u=v=1/2$,
$\gamma=0$, a neutral eigenvalue $\lambda_1=-\frac{2}{3}k^2+\rmO(k^4)$,
with eigenvector $(1+\frac{2}{3}\rmi k,1-\frac{2}{3}\rmi k)^T+\rmO(k^2)$,
and therefore chooses an ansatz
\[
u(t,x)=\eps A(\eps x,\eps^2 t)+\frac{2}{3}\eps^2 A_X(\eps x,\eps^2 t),\quad
v(t,x)=\eps A(\eps x,\eps^2 t)-\frac{2}{3}\eps^2 A_X(\eps x,\eps^2 t).
\]
We denote the variables corresponding to the diffusive scaling by
$T=\eps^2 t,X=\eps x$. Then substituting into the equation, expanding in $\eps$,
and projecting onto the kernel, we find at order $\eps^3$
\[
A_T=\frac{2}{3}A_{XX},
\]
a simple diffusion equation. 

We emphasize that the diffusive effect is generated by the interaction of
tumbling and transport, without explicit diffusion in the system! This can be
understood somewhat more directly in the interaction of tumbling and
transport on the linear level, deriving a damped wave equation for the total 
mass $q=u+v$, as described in the introduction \eqref{e:kac}.  A description of dynamics in the
unstable or marginally stable regime,  then appears to be difficult for a 
variety of reasons. First,  since all wavenumbers are simultaneously unstable, 
an expansion around one particular mode is unlikely to capture dynamics. 
Moreover, expansions around the most unstable mode $k=\infty$ appear 
challenging. Lastly, resulting patterns are not of small amplitude against 
the uniform unstable background, such that  small amplitude expansions will 
likely not capture relevant phenomena.

\paragraph{Dependence on tumbling rates.}

Our results are for a very specific tumbling rate, only, and one can easily 
argue for different rates. On the other hand, the linear selection results and 
the existence of traveling waves depend on the geometric shape of the curves 
of equilibria, only. From this perspective, it turns out that the class of 
nonlinearities represented by our specific choice is quite a bit larger. 

\begin{figure}
\includegraphics[height=1.5in]{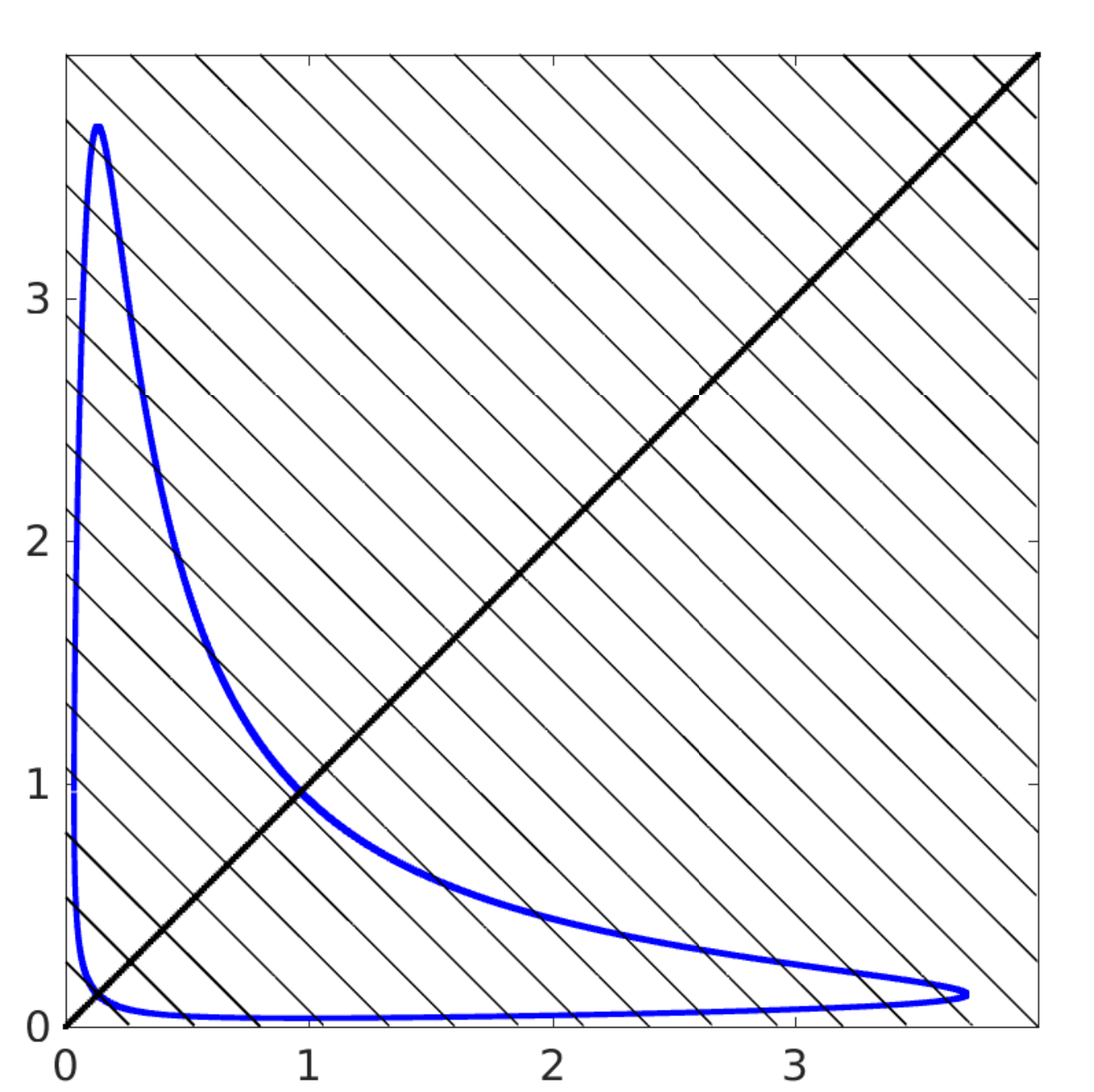}\hfill
\includegraphics[height=1.5in]{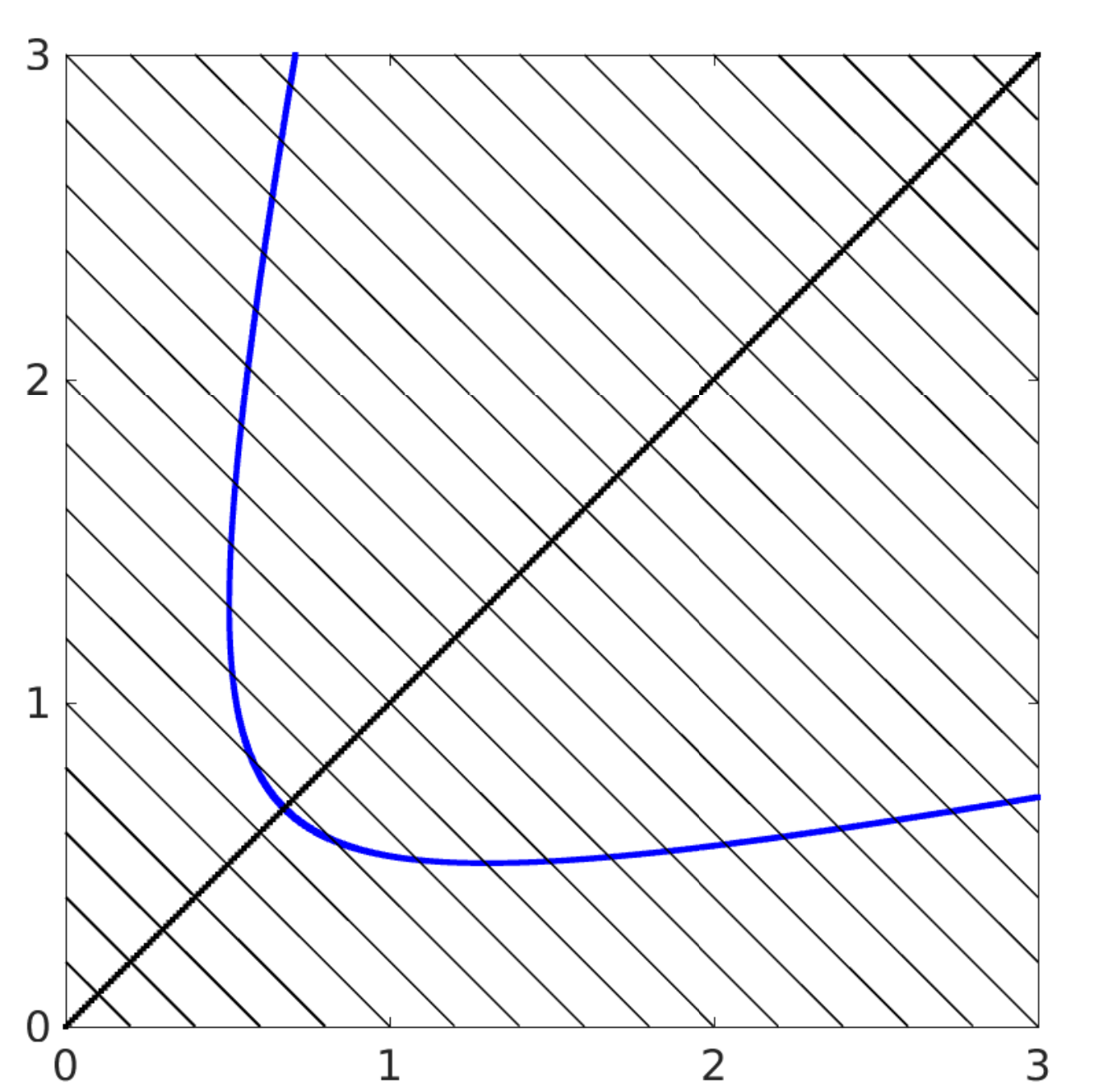}\hfill
\includegraphics[height=1.5in]{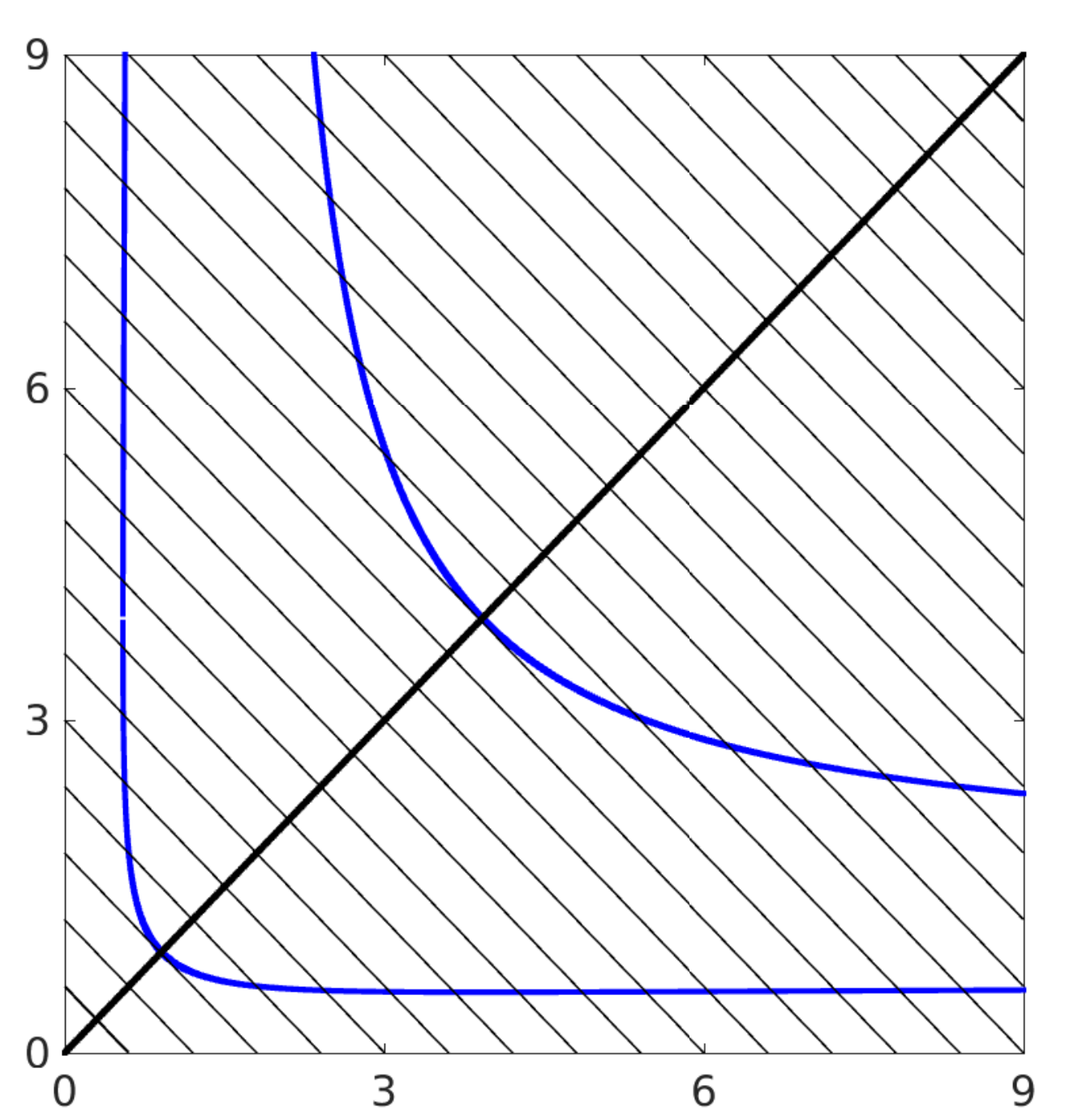}\hfill
\includegraphics[height=1.435in]{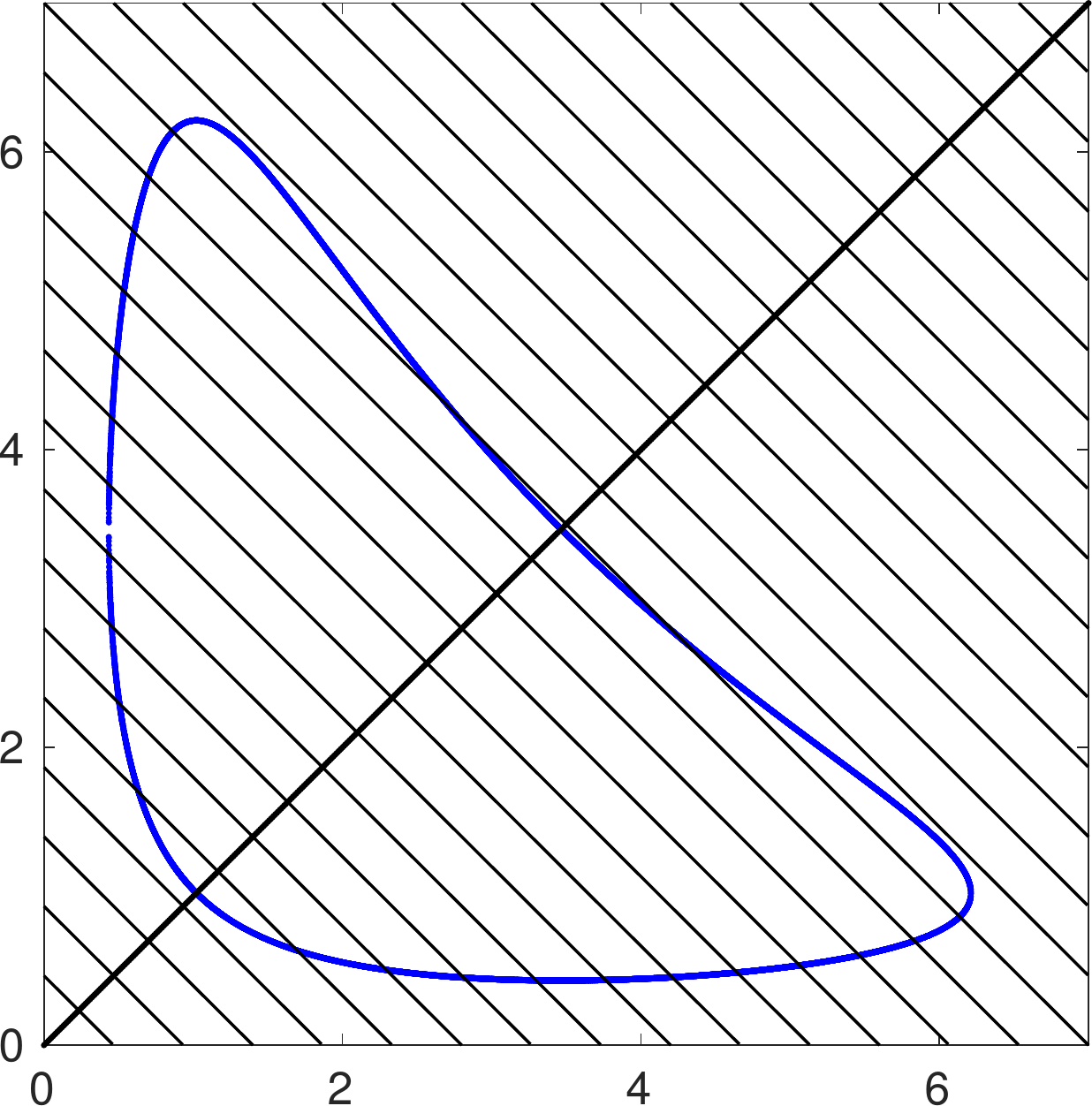}
\caption{Equilibria for different tumbling rates: $g(v)=\mu+\frac{v^2}{1+\gamma v^2}$ with $\mu=0.166,\gamma=1$ (left); $g(v)=\mu+\frac{v^2}{1+\gamma (u+v)^2}$, $\mu=0.2,\gamma=0.7$ (center left); $g(v)=\mu+
\frac{ v^3}{1+\gamma v^2}$, $\mu=0.45,\gamma=1$ (center right); $g(v)=\mu+
\frac{ v^2}{1+\gamma v^3}$, $\mu=1,\gamma=0.01$ (right).  }\label{f:tro}
\end{figure}

As a first generalization, one could introduce add a parameter in
 front of the linear part of the tumbling rate, $g(v)=\mu+\frac{v^2}{1+\gamma v^2}$.
Figure 
\ref{f:tro} shows that equilibrium configurations are qualitatively similar 
to the case $\mu=1$, studied here. 

Interestingly, the structure of the set of asymmetric equilibria changes 
qualitatively when nonlinear saturation of tumbling rates is caused by 
increase of \emph{total mass} in the system, 
$g(v)=\mu+\frac{v^2}{1+\gamma (u+v)^2}$, cf. \cite{LutscherStevens, 
PrimiStevensVelazquez}. Such nonlinearities incorporate a slightly different 
sensing of also the total population as opposed to only sensing  
the opposite moving 
bacteria. One notices that for such systems, the branch of asymmetric 
equilibria does not re-destabilize; see Figure \ref{f:tro}. Generally, 
asymmetric equilibria bifurcate as PDE-unstable branches that subsequently 
stabilize; the resulting stable branch extends to infinity. Moreover, the 
asymmetric branch appears in a bifurcation from infinity, rather than through 
the spontaneous emergence of a sub- and a super-critical pitchfork bifurcation 
at finite mass.
Following \cite{LutscherStevens}, one can generalize further by allowing 
different power laws for small and large $v$,  $g(v)=\mu+
\frac{ v^p}{1+\gamma v^q}$. A priori bounds based on the comparison 
principle require $p\leq q$ \cite{LutscherStevens}.

We notice that for $p>q$, asymmetric branches of equilibria extend to infinity; see Figure \ref{f:tro}. 
{It would be nice to further explore for which parameter regimes
solutions blow up in finite time, i.e. reflecting the initiation 
of fruiting body formation as discussed for the model in \cite{LutscherStevens}.
Ideally, only small parameter changes in the model would result in
the formation of ripples, of aggregates or in the initiation of self-organization.} 
\begin{figure}
\includegraphics[height=1.2in]{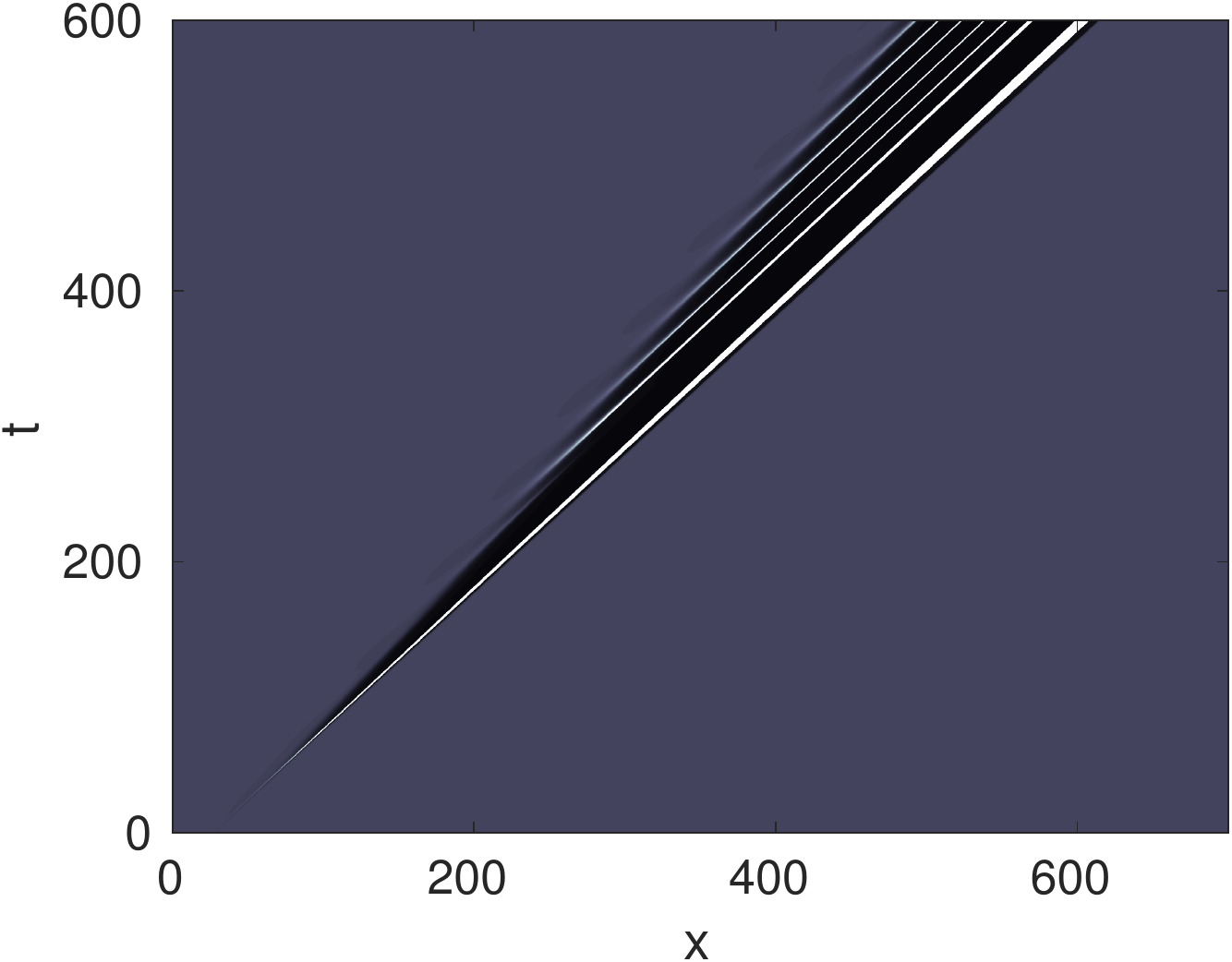}\hfill
\includegraphics[height=1.2in]{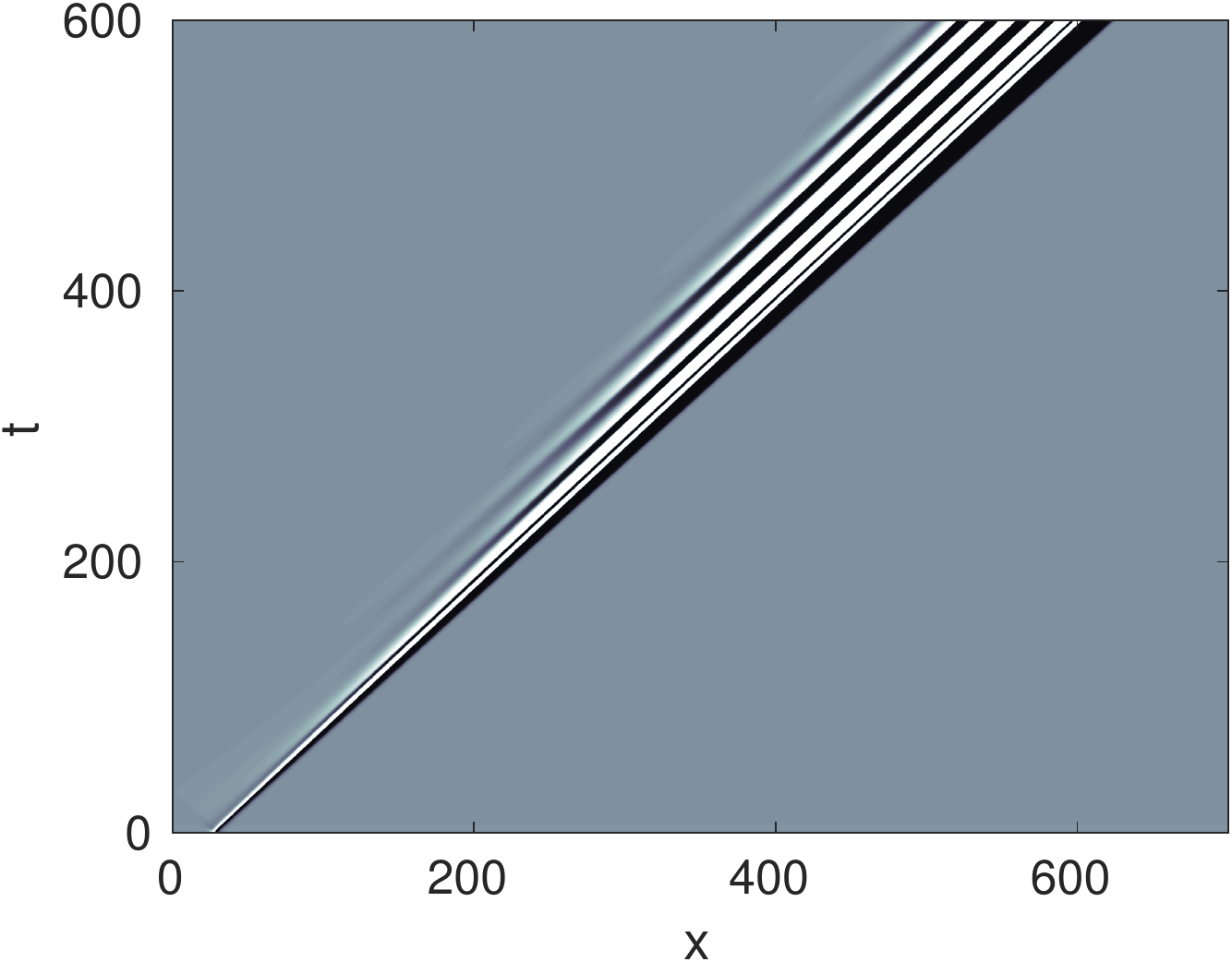}\hfill
\includegraphics[height=1.2in]{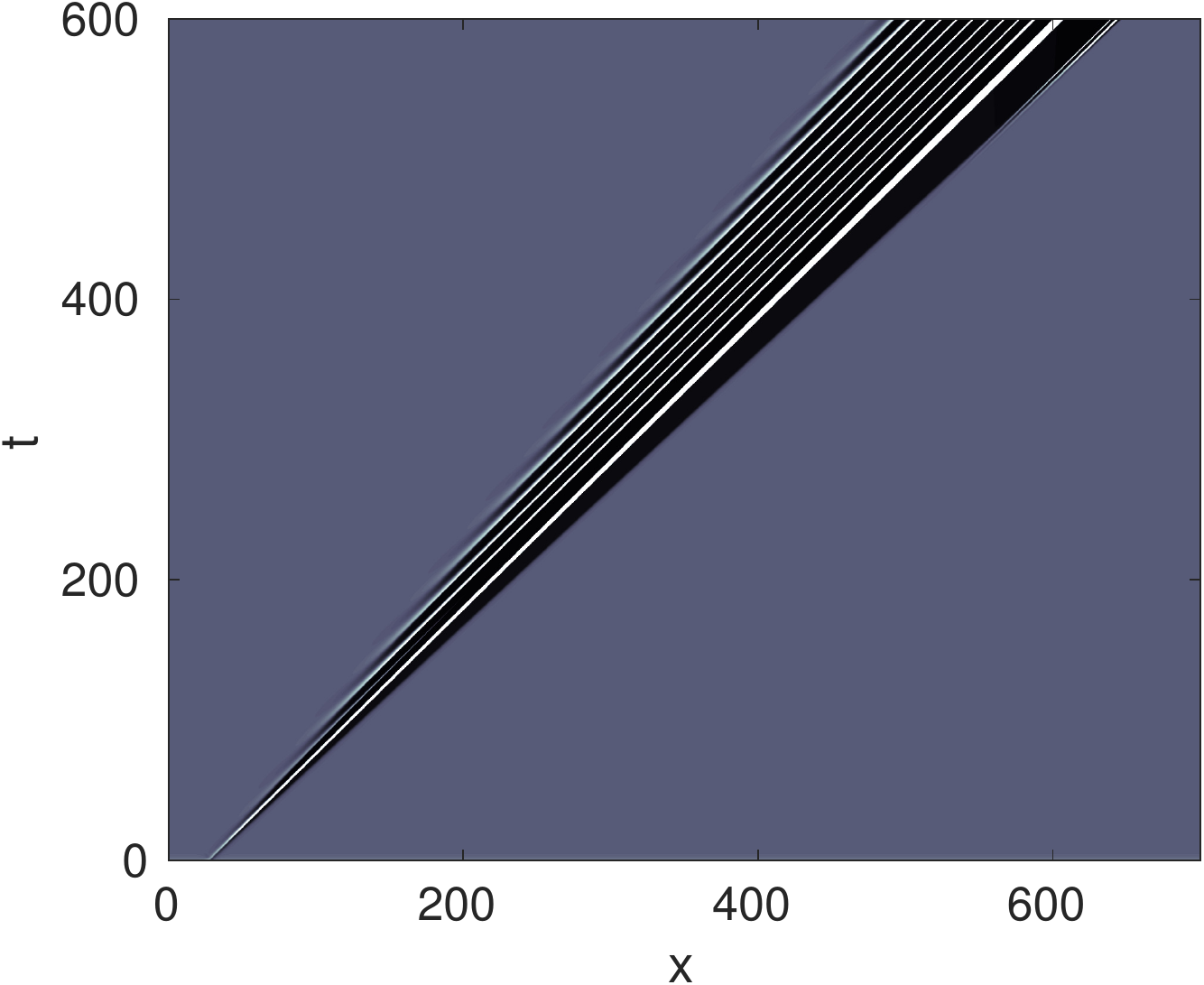}\hfill
\includegraphics[height=1.2in]{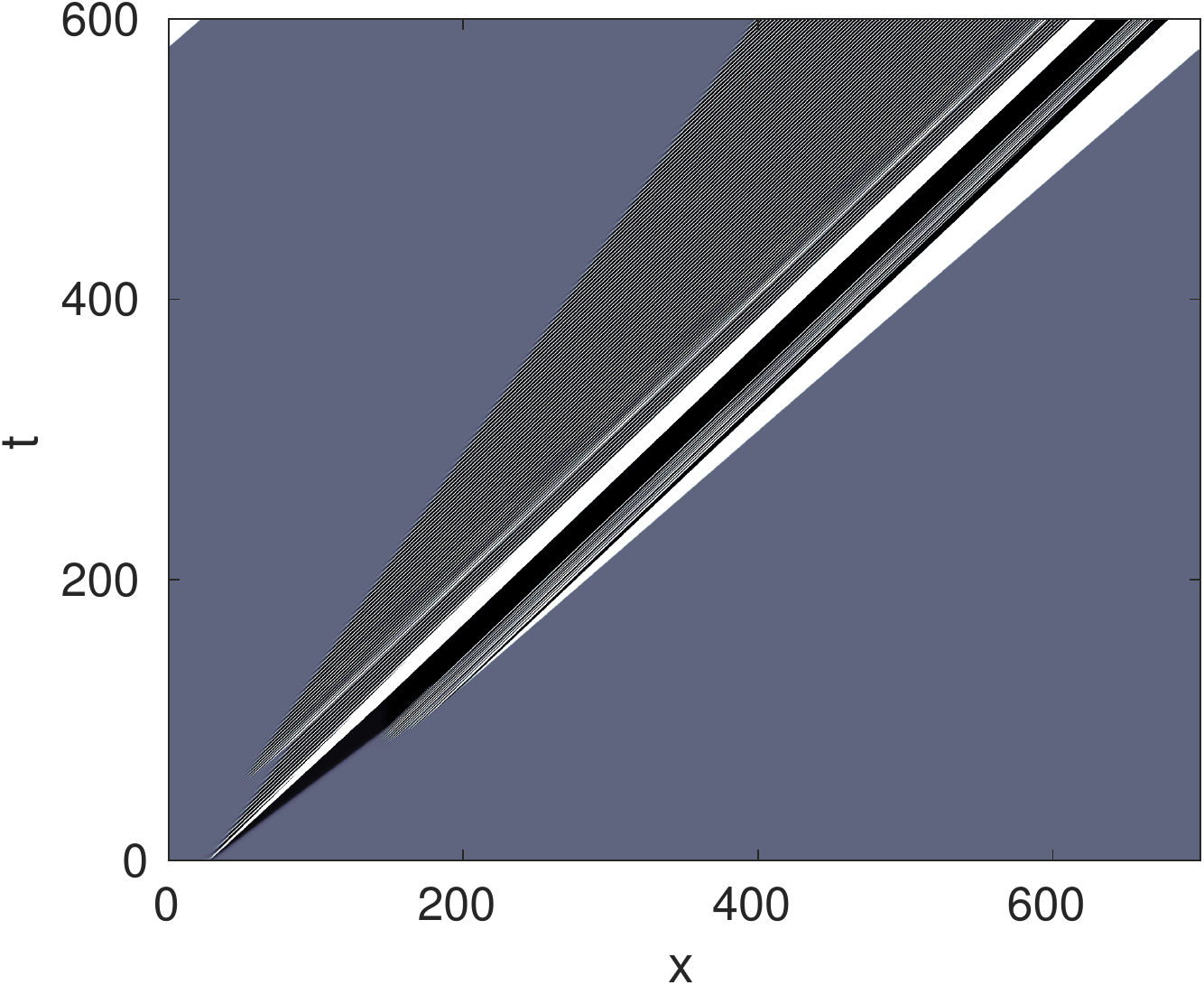}
\caption{Space-time plots of the $v$-component resulting from perturbations of asymmetric equilibria through localized perturbations for tumbling kinetics as in Figure \ref{f:tro} (from left to right). Initial mass values were $u=0.06,v=0.3$ (left, $k\sim 0.37$), $u=0.35,v=1 $ (center left, $k\sim 0.5$), $u=0.3,v=2$ (center right $k\sim 0.57$), and $u=0.55,v=2$ (right,  $k\sim 2.1$).  }\label{f:tro2}
\end{figure}

In direct simulations, we found little differences when starting with localized perturbations of asymmetric states; see Figure \ref{f:tro2}.  We did not see patterning when perturbing symmetric equilibria in the case of saturation by total mass.

\paragraph{Coherent structures and stability.}
Striving to put the present analysis on a stronger theoretical footing, one would want to analyze coherent structures and their stability. As a first step, one would study the stability of piecewise constant wave trains using Floquet-Bloch theory. In a different direction, it would be interesting to analyze the effect of small diffusion on wave trains, fronts, and interfaces, 
possibly including dispersive terms $-\partial_{xxx}$ or even higher-order 
diffusion $\partial_{xxxx}$. Numerical observations from higher-order upwind 
finite-difference simulations suggest that stability properties of  the 
jump interface can change in subtle ways under diffusive or dispersive regularization. Ultimately, the phenomena described here should of course be understood in terms of invasion fronts, their existence, and their stability.

\paragraph{Wavenumber selection through growth elsewhere.}

The phenomena observed here bear a striking resemblance to wavenumber 
selection in models for recurrent precipitation \cite{gms,diffode},
\begin{align*}
u_t&=u_{xx}-g(v)-u\\
v_t&=g(v)+u,
\end{align*}
where $g(v)$ is non-monotone, for instance $g(v)=v(1-v)(v-a)$, $a\in (0,1)$. 
White noise perturbations of unstable spatially homogeneous equilibria result 
in spatially decorrelated patterns of arbitrarily fine spatial scale, while 
shot noise perturbations give rise to patterns with a dominant nonzero spatial 
wavenumber. Structurally, the system is similar to our system as it possesses 
a reflection symmetry (which however does not involve the dependent variables) 
and a conservation law for the total mass. We also have a plethora of spatially periodic patterns, $u(x)=0,v(x)\in\{0,1\}$, analogous to the traveling waves described in Section \ref{s:3}, but independent of time $t$. 

Spatially constant equilibria with $g'(v)>0$ are unstable but fastest growing wavenumbers are either $0$ or $\infty$ as observed in our models. Again, invasion processes do select finite, nonzero wavenumbers and 
observations in direct numerical simulations  agree quite well with linear 
predictions, 
obtained in a similar fashion to our analysis in Section \ref{s:4}. Furthermore, selected patterns are discontinuous, stable, yet subject to coarsening when a small diffusivity is added in the second equation. 

A more subtle similarity is the ``instantaneous coarsening'' observed here: a wave maximum nucleating at the leading edge of the growth process may not develop into a large amplitude peak of a wave train, or merge with a previous peak, as can be seen in 
{ Figure \ref{f:5}, for the first two peaks or the 13th peak, respectively}. This mechanism was demonstrated numerically in \cite{gms,diffode} and associated with resonances. A more detailed study can be found in \cite{ch} for the Cahn-Hilliard equation. 
%


%
%
%
%

\section*{Appendix}

We outline a basic functional analytic setting that allows us to show local existence and uniqueness of solutions, as well as smooth dependence on initial data. { Previous work in the literature does not quite cover the setting of bounded yet discontinuous functions that we are interested in, here; see for instance \cite{Hillen10} for a linear theory, including boundary conditions.} We therefore solve 
\[
u_t=u_x+f(u,v),\qquad v_t=-v_x-f(u,v),
\]
with initial data $u_0(x),v_0(x)\in X$, a Banach space, using the variation-of-constants formula
\begin{align}
u(t,x)&=u(0,x+t)+ \int_0^t f\left(u(\tau,x+(t-\tau)),v(\tau,x+(t-\tau))\right)\rmd \tau\\
v(t,x)&=v(0,x-t)+ \int_0^t f\left(u(\tau,x-(t-\tau)),v(\tau,x-(t-\tau))\right)\rmd \tau.
\end{align}
For $t\in [-\delta,\delta]$, sufficiently small, this system of integral equations defines a contraction mapping principle and yields local existence of solutions in a similar way as the Picard-Lindel\"of iteration, provided that $X$ controls supremum norms, that is, choosing for instance $X=BC^0, BC^0_\mathrm{unif}, L^\infty,$ or $X=BV$. Note that solutions will not necessarily be continuous in time for $X=L^\infty$. Mimicking the existence proof for ODEs further, we also obtain continuous dependence on initial data in these spaces for fixed times $t\neq 0$.  

From the integral formulation, it is immediately clear that solutions where $f\equiv 0$ are given through simple right- and left shifts of $u$ and $v$, respectively.

%

%


\end{document}